\documentclass[aip,pop,reprint,amsmath,amssymb,graphicx,floatfix]{revtex4-1}
\usepackage{dcolumn}
\usepackage{bm}
\usepackage{graphics, graphicx,wrapfig}
\usepackage{color}
\usepackage{soul}

\newcommand{\dmytro}[1]{{\color{black}#1}}
\newcommand{\matthias}[1]{{\color{black}#1}}

\begin{document}

\title{Numerical study of tearing mode seeding in tokamak X-point plasma}

\author{Dmytro Meshcheriakov}
	\email[]{dmytro.meshcheriakov@ipp.mpg.de}
	\affiliation{Max Planck Institute for Plasma Physics, Boltzmannstr.~2, 85748 Garching b.M., Germany}
	\affiliation{Max-Planck/Princeton Research Center for Plasma Physics}
\author{Matthias Hoelzl}
	\affiliation{Max Planck Institute for Plasma Physics, Boltzmannstr.~2, 85748 Garching b.M., Germany}
\author{Valentin Igochine}
	\affiliation{Max Planck Institute for Plasma Physics, Boltzmannstr.~2, 85748 Garching b.M., Germany}
\author{Sina Fietz}
	\affiliation{Max Planck Institute for Plasma Physics, Boltzmannstr.~2, 85748 Garching b.M., Germany}
\author{Francois Orain}
	\affiliation{Max Planck Institute for Plasma Physics, Boltzmannstr.~2, 85748 Garching b.M., Germany}
	\affiliation{CPHT, Ecole Polytechnique, 91128 Palaiseau cedex, France}
\author{Guido T.A. Huijsmans}
    \affiliation{CEA Cadarache, IRFM, 13108 St. Paul Lez Durance Cedex, France}
    \affiliation{Technische Universiteit Eindhoven, P.O. Box 513, 5600 MB Eindhoven, Netherlands}
\author{Marc Maraschek}
	\affiliation{Max Planck Institute for Plasma Physics, Boltzmannstr.~2, 85748 Garching b.M., Germany}
\author{Mike Dunne}
	\affiliation{Max Planck Institute for Plasma Physics, Boltzmannstr.~2, 85748 Garching b.M., Germany}
\author{Rachel McDermott}
	\affiliation{Max Planck Institute for Plasma Physics, Boltzmannstr.~2, 85748 Garching b.M., Germany}
\author{Hartmut Zohm}
	\affiliation{Max Planck Institute for Plasma Physics, Boltzmannstr.~2, 85748 Garching b.M., Germany}
\author{Karl Lackner}
	\affiliation{Max Planck Institute for Plasma Physics, Boltzmannstr.~2, 85748 Garching b.M., Germany}
\author{Sibylle G\"unter}
	\affiliation{Max Planck Institute for Plasma Physics, Boltzmannstr.~2, 85748 Garching b.M., Germany}
\author{ASDEX Upgrade Team}
	\affiliation{Max Planck Institute for Plasma Physics, Boltzmannstr.~2, 85748 Garching b.M., Germany}
\author{{EURO}fusion {MST1} Team}
    \affiliation{See author list of {MST1} Team in H. Meyer et al., Nucl. Fusion 57, 102014 (2017)}

\date{Version of \today}

\begin{abstract}
%Tearing modes (TMs) forming magnetic islands inside the core plasma degrade plasma confinement and are important cause for disruptions in tokamaks\cite{Schuller1995}. Although TMs are linearly stable, in most cases they are frequently observed in present experiments since a sufficiently large seed perturbation can drive TMs into a regime where they become non-linearly unstable, grow further independently of the seed perturbation and remain present also after the seed has decayed away. Such neoclassical tearing modes (NTMs) are driven by the helical perturbation of the bootstrap current~\cite{LaHaye2006} resulting from pressure flattening inside the seed island~\cite{Fitzpatrick1995}. Seed perturbations are often provided by other plasma instabilities~\cite{Gude1999} like sawtooth crashes at the plasma center.
\dmytro{
A detailed understanding of island seeding is crucial to avoid (N)TMs and their negative consequences like confinement degradation and disruptions. In the present work, we investigate the growth of $2/1$ islands in response to magnetic perturbations. Although we use externally applied perturbations produced by resonant magnetic perturbation (RMP) coils for this study, results are directly transferable to island seeding by other MHD instabilities creating a resonant magnetic field component at the rational surface. Experimental results for $2/1$ island penetration from ASDEX Upgrade are presented extending previous studies. Simulations are based on an ASDEX Upgrade L-mode discharge with low collisionality and active RMP coils. Our numerical studies are performed with the 3D, two fluid, non-linear MHD code JOREK. 
All three phases of mode seeding observed in the experiment are also seen in the simulations: first a weak response phase characterized by large perpendicular electron flow velocities followed by a fast growth of the magnetic island size accompanied by a reduction of the perpendicular electron velocity, and finally the saturation to a fully formed island state with perpendicular electron velocity close to zero. Thresholds for mode penetration are observed in the plasma rotation as well as in the RMP coil current. A hysteresis of the island size and electron perpendicular velocity is observed between the ramping up and down of the RMP amplitude consistent with an analytically predicted bifurcation. The transition from dominant kink/bending to tearing parity during the penetration is investigated.}
\end{abstract}


\maketitle 

% ************************************
\section{Introduction}
% ************************************

Physics of forced magnetic reconnection in magnetically confined plasmas is crucial to understand magnetic island formation and associated degradation of the plasma confinement and potentially a disruption of the plasma. An instability associated to magnetic reconnection develops in the presence of finite plasma resistivity or other non-ideal effects and is driven by both the equilibrium current density gradient (classical tearing mode) and a "hole" in the bootstrap current profile (neoclassical tearing mode or NTM)~\cite{Biskamp1993,Chang1998,Smolyakov1993,Zohm2001} caused by a flattening of the temperature distribution inside the magnetic island~\cite{Fitzpatrick1995}.

It was shown both theoretically~\cite{Drake1983, Scott1985,Scott1987,Waelbroeck1989,Smolyakov1993,Meshcheriakov2012} and experimentally~\cite{Gude1999,Zohm2001,Sauter2002,LaHaye2006,Igochine_PoP2014,Igochine_NF2017} that NTMs are linearly stable and require a seed magnetic island for their growth which is provided by triggers like other MHD instabilities. Resistive MHD predicts tearing modes to be linearly unstable when the parameter $\Delta^\prime$, measuring the available magnetic free energy, is positive. Two-fluid pressure gradient effects, the ion polarization current and the toroidal curvature, so-called Glasser-Greene-Johnson effect, provide additional stabilizing effects\cite{Ara,Waelbrock_PRL2001,GGJ1976}.
When a tearing mode is linearly stable, a sufficiently large initial seed island can lead to further island growth since it causes a helical perturbation of the temperature distribution and consequently the bootstrap current, which acts destabilizing. Nonlinear effects and toroidal mode coupling enable the generation of a seed island from an initial perturbation with a different helicity. For example, sawtooth post-cursors with helicity $m/n = 1/1$ were observed~\cite{Gude1999,Igochine_PoP2014} to produce a $2/1$ component acting as seed for a $2/1$ magnetic island. \\

A basic theoretical framework of tearing mode interactions with a static external magnetic perturbation in cylindrical geometry is proposed by Fitzpatrick~\cite{Fitzpatrick1993}. In this work, externally applied MPs are treated as modified edge boundary conditions. The interaction of the external MPs with the helical perturbation current associated to a magnetic island results in the modification of the island width evolution and a rise of a local $\mathbf{j}\times \mathbf{B}$ torque in the vicinity of the island. If the island frequency deviates from its natural frequency, the plasma exerts a viscous restoring torque onto the island. The general non-linear tearing mode stability problem is then treated as a balance of  the plasma inertia in the island and the sum  of the local electromagnetic and viscous torques. In the presence of perpendicular electron velocity, static RMPs in the laboratory frame correspond to time varying RMPs in the electron fluid frame, and therefore induce a current hindering their penetration~\cite{NardonNF2010,OrainPoP2013}. 

\begin{figure}[htbp]
	\centering
	\includegraphics[width=0.8\linewidth]{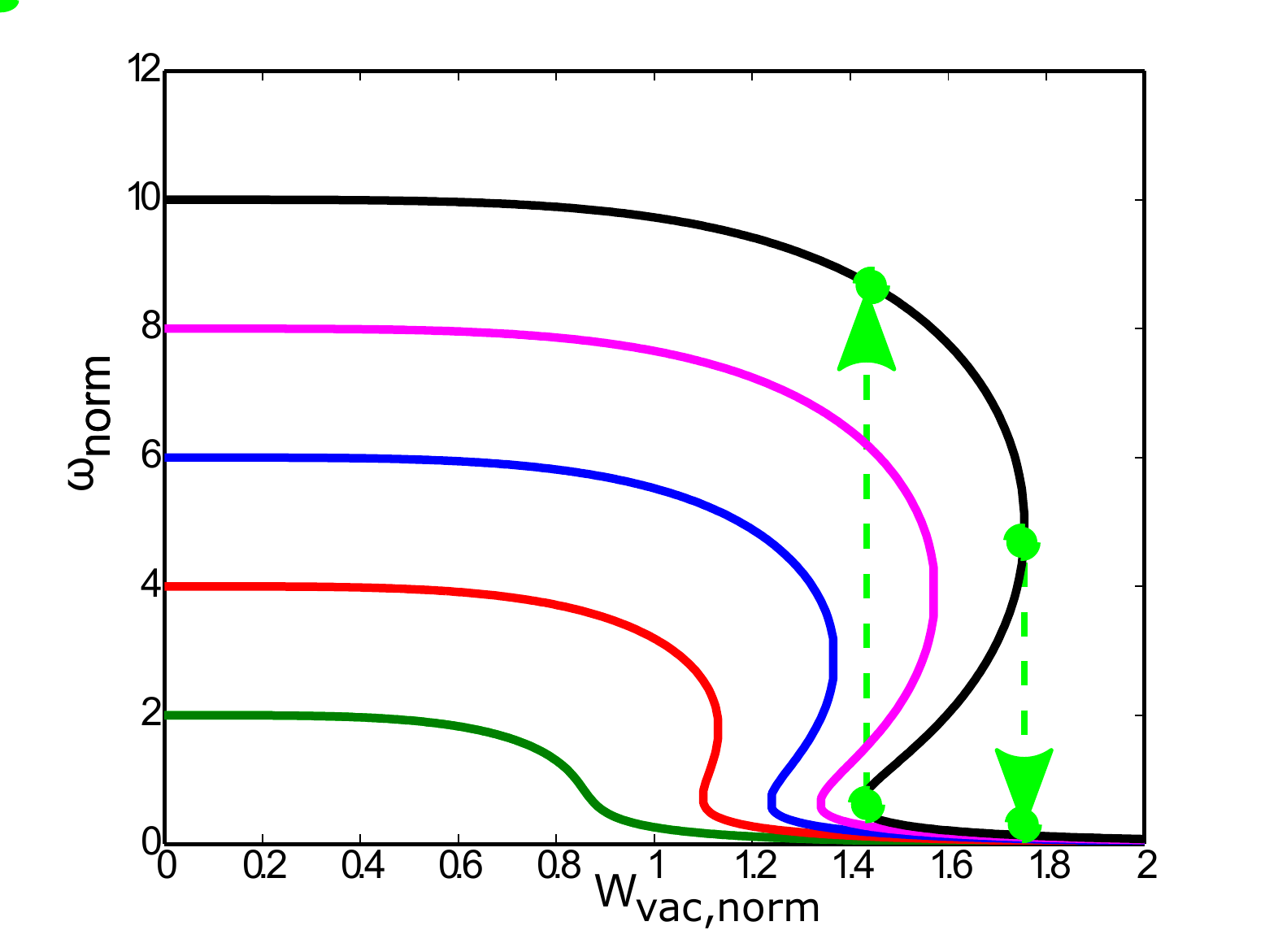}
	\caption{The steady state mode rotation frequency is plotted versus the applied perturbation amplitude according to analytical theory (see Ref.~\cite{Fitzpatrick1993}, also for the normalization). The different curves correspond to different background rotation values, i.e., different rotation frequencies in the absence of the perturbation. At high rotation frequencies, a bifurcation is observed, where the force balance does not have a unique solution any more. Finally, when the rotation frequency has dropped by a factor of about two, the plasma can undergo a fast transition to the penetrated state with a rotation frequency close to zero. A hysteresis of rotation frequency (and correspondingly penetrated island size) is expected between a ramping up and down of the perturbation amplitude except for low background rotation where the solution remains unique for all perturbation amplitudes.}
	\label{om_Wvac_om0}
\end{figure}

A steady state is obtained when the viscous torque is balanced by the electromagnetic one. Solutions corresponding to different values of initial plasma rotation (and, as a result, island natural frequency) are shown in Fig.~\ref{om_Wvac_om0}. For large initial plasma rotation, a bifurcation is observed and the plasma undergoes a sudden transition at about half the initial rotation frequency to a non-linear island state, characterized by low plasma rotation and a large magnetic island.  This transition is generally called mode penetration. Due to the bifurcation, a hysteresis is expected for the back transition: The island remains on the lower branch of the figure in the region where the force balance equation does not have a unique solution.\\

In the present paper, the seeding of a tearing mode by externally applied magnetic perturbations is studied in presence of realistic poloidal and toroidal background rotation. The successive mode evolution and the impact on confinement are also addressed. These questions are investigated with the toroidal nonlinear MHD code JOREK~\cite{Huysmans2007,Czarny2008}, which includes anisotropic heat transport, two-fluid diamagnetic effects\cite{Orain2013}, neoclassical friction, and toroidal rotation in realistic tokamak X-point geometry. A discharge in low density L-mode plasmas~\cite{Fietz_EPS2015} in the ASDEX Upgrade~\cite{Kallenbach2017} tokamak (AUG\#30734) was chosen as basis for our studies. \dmytro{Bootstrap current drive is not considered in the present work since we are predominantly interested in the seeding respectively mode penetration, not the further non-linear evolution. Also the effect of neoclassical toroidal viscosity (NTV), which would enhance mode penetration is not taken into account in the present paper. Inclusion of NTV and bootstrap current as well as quantitative comparisons to the experiment are left for future studies.}\\

This paper is organized as follows. In section~\ref{exp}, we briefly show experimental observations from ASDEX Upgrade. Section~\ref{simresults} introduces the JOREK code, the simulation setup, and results of our simulations of $2/1$ mode penetration in ASDEX Upgrade reproducing qualitatively all experimental observations and analytical predictions. This includes observations of penetration thresholds in coil current and background rotation velocity, a hysteresis between ramp-up and ramp-down and a transition from kink- to tearing parity at the resonant surface. Summary and conclusions are given in section~\ref{summary}.

% ************************************
\section{\label{exp}Mode penetration in ASDEX Upgrade experiments}
% ************************************

\matthias{Experimental results are shown in this section extending previous work described in Ref.~\cite{Fietz2015}.}
Here we refer to the ASDEX Upgrade discharge number $\#30734$. Three distinguished phases were observed in this experiment while the current in the MP field coils with the dominant mode number $n=1$ was slowly ramped up (Figure~\ref{Wvst})): 

\begin{figure}[htbp]
	\centering
	\includegraphics[width=0.8\linewidth]{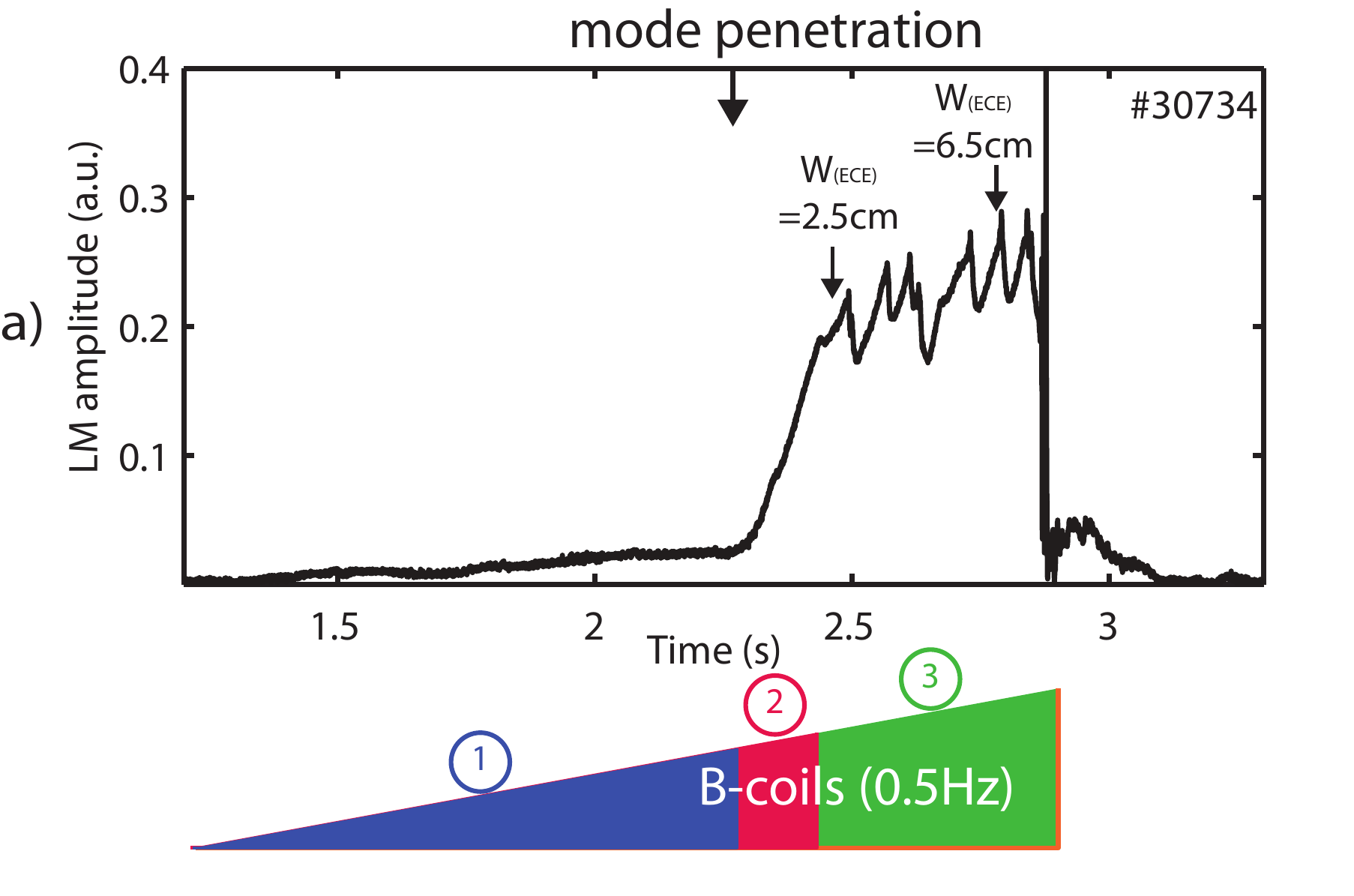}
	\caption{Amplitude of the $n=1$ magnetic field perturbation measured by the locked mode detection system and evolution of the current in the B-coils (below). The weak response phase (1), the penetration phase (2), and the saturation phase (3) are clearly visible.}
	\label{Wvst}
\end{figure}

\begin{figure}[htbp]
	\centering
    \includegraphics[width=0.8\linewidth]{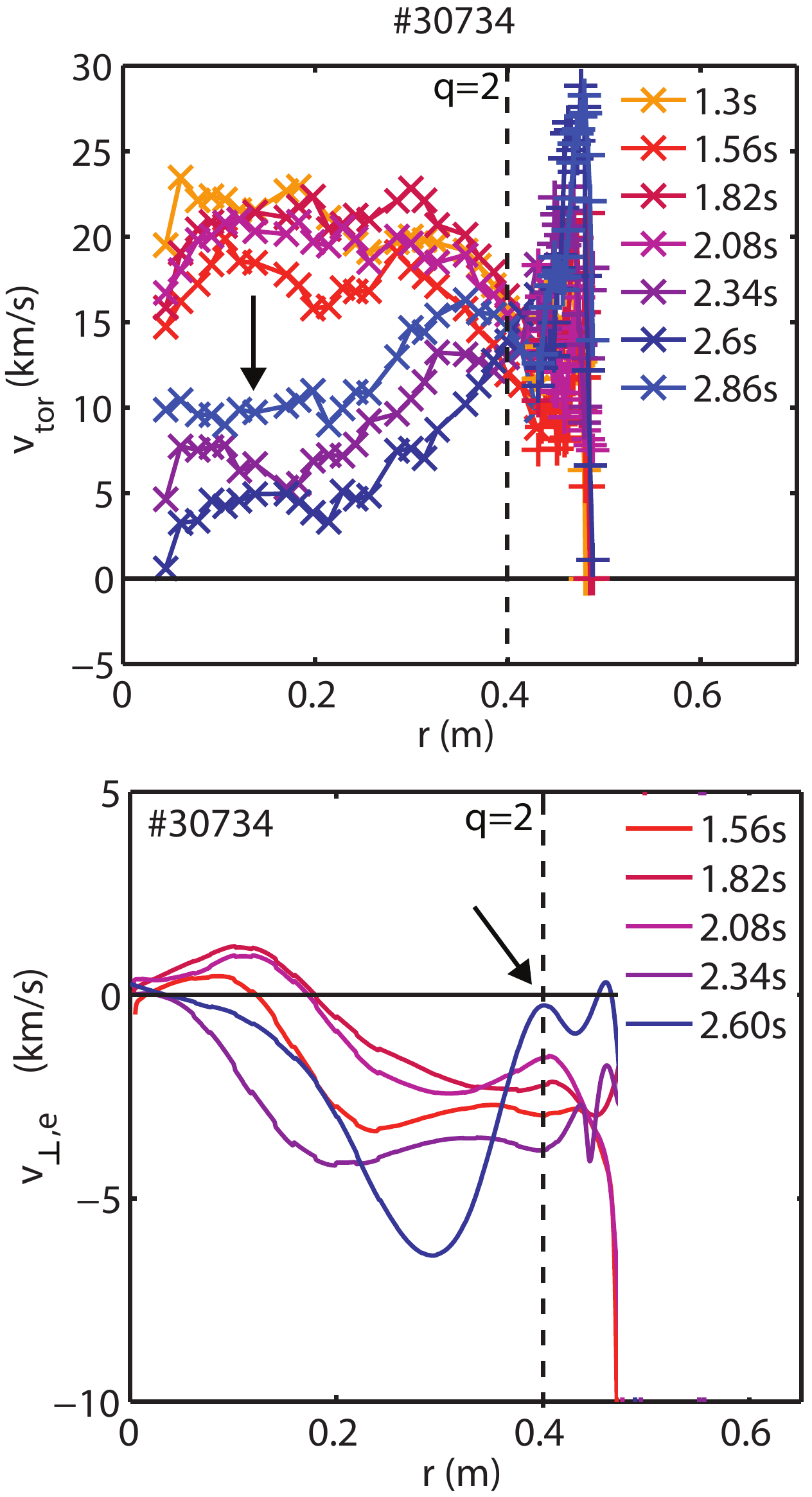}	
	\caption{Experimental toroidal rotation~(upper) and perpendicular electron velocity~(lower) profiles measured by charge exchange recombination spectroscopy\cite{Viezzer2012} at several time points of the experiment shown in Figure~\ref{Wvst}. The toroidal rotation velocity is reduced after the penetration in the plasma core. % and enhanced towards the plasma edge.
	The perpendicular electron velocity drops to zero at the $q=2$ surface when the $2/1$ mode is fully penetrated. Fast penetration approximately sets in when the original perpendicular electron velocity has dropped by a factor of two consistent with the analytical predictions described above.}
	\label{V_exp}
\end{figure}
1
In the first phase, denoted weak plasma response phase, the plasma response follows the amplitude of the magnetic perturbation  approximately linearly. In this phase, screening is strong and the residual perturbation on the resonant surface is not sufficient to drive magnetic reconnection. In the second phase, the perturbation exceeds a certain threshold and becomes strong enough to slow down the rotation strong enough such that the transition point is reached and forced reconnection takes place at the q=2 surface. The resulting $(2/1)$ magnetic island is observed in the magnetic data and in the electron temperature. In the third phase, the island growth slows down and is interrupted by some minor disruptions.

\begin{figure}[htbp]
	\centering
    \includegraphics[width=0.8\linewidth]{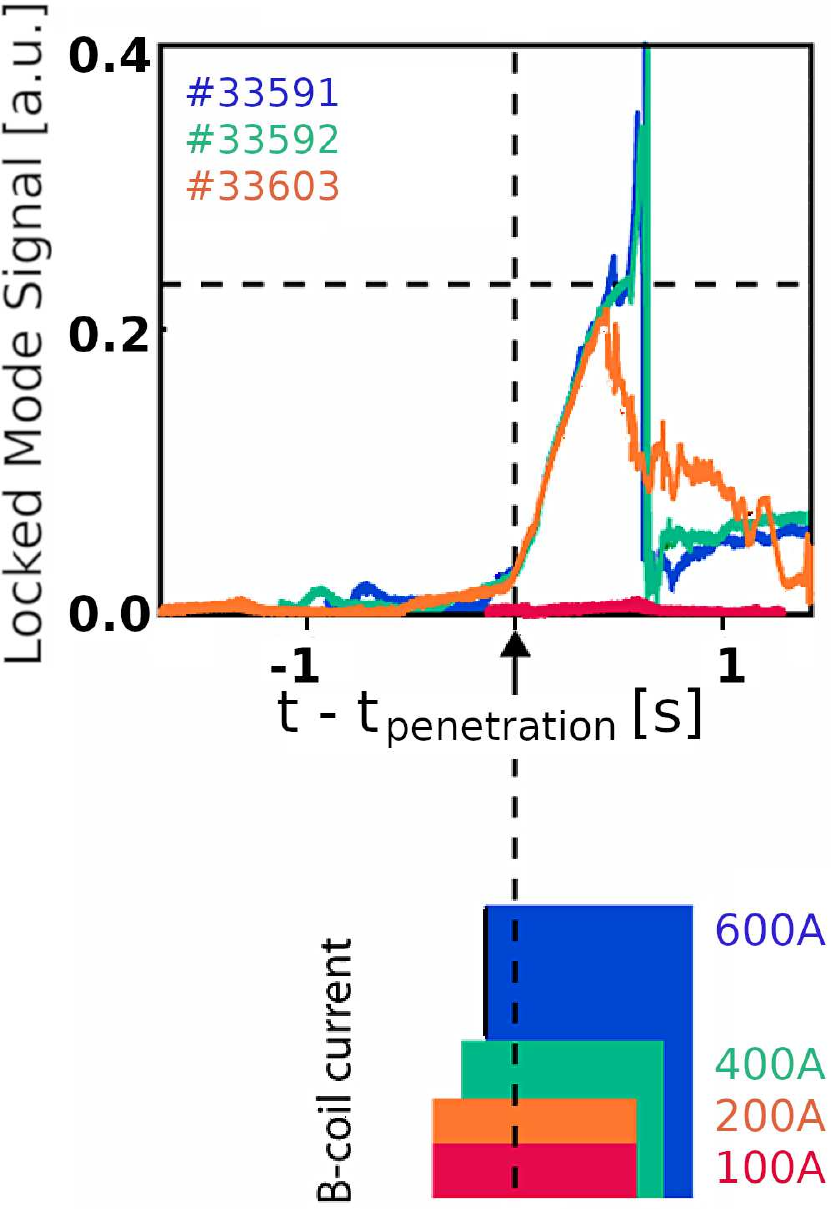}
    \includegraphics[width=0.8\linewidth]{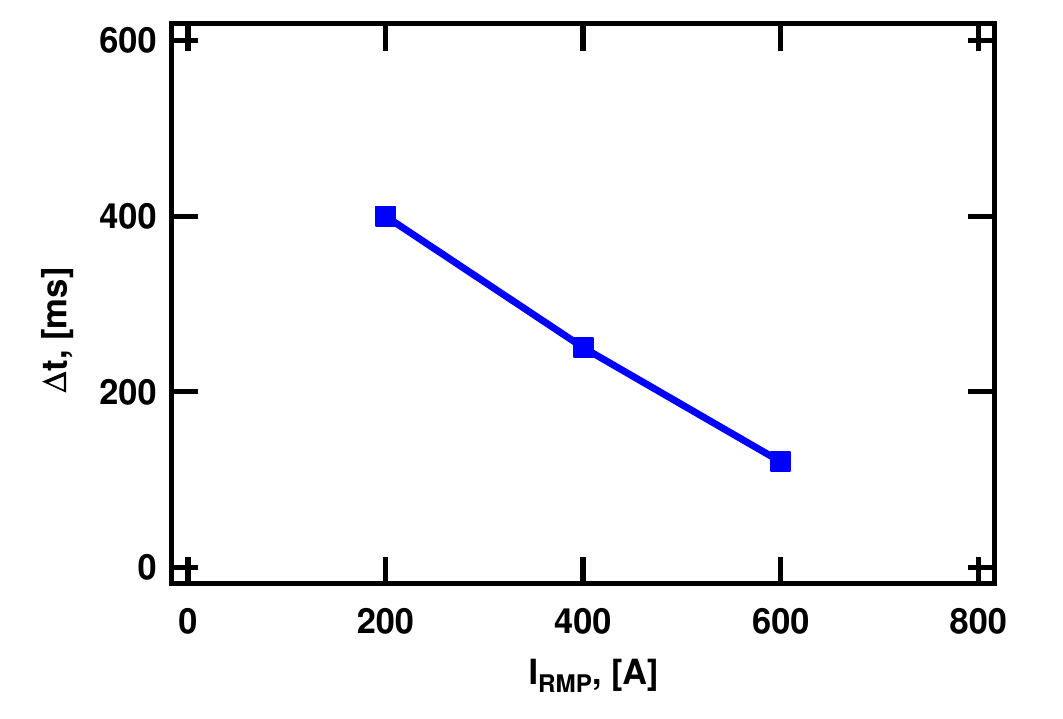}
	\caption{Dependence of the mode penetration delay on the RMP current amplitude. For larger perturbation amplitudes, the mode penetration takes place faster. Delays are approximately 120 ms for 600A, 250 ms for 400A and 400 ms for 200A. At a coil current of 100A, no mode penetration is observed at all.}
	\label{delay_exp}
\end{figure}

During the first phase (Figure~\ref{V_exp}), the core toroidal rotation~\cite{McDermottNF2014} decreases up to the point of mode penetration. A fast drop is observed in the second phase and the value remains almost constant during the whole third phase. %Contrary, the edge toroidal rotation is observed to increase. 
%The perpendicular electron velocity profile was calculated using the experimentally measured toroidal rotation (from the charge-exchange recombination spectroscopy) and electron temperature profiles (from electron cyclotron emission spectroscopy).
\matthias{The perpendicular electron velocity is calculated from the measured ExB velocity and the electron diamagnetic drift velocity. The ExB velocity is evaluated from the Er profile measured with charge exchange recombination spectroscopy via the radial force balance equation, while the electron diamagnetic drift velocity is calculated from the measured electron temperature and density profile. In the plasma core, the toroidal rotation velocity is the dominant term in the radial force balance equation and hence, we have used this term in the radial force balance equation to evaluate the ExB velocity. Further information about the measurements in ASDEX Upgrade is found for instance in Ref.~\cite{Viezzer_2013}.}
In the first phase, the motion of the electron fluid across the field lines at the resonant surfaces screens the RMPs hindering their penetration \matthias{in agreement with previous findings that the electron perpendicular rotation is a key factor for the screening of magnetic perturbations by the plasma~\cite{Waelbroeck2003,Nardon2007,Heyn_2008,Orain2013}}. Mode penetration corresponds to a drop of the perpendicular electron velocity to approximately zero. These experiments confirm the predicted slow decrease of the plasma rotation towards the time of mode penetration and the small electron perpendicular velocity when an island is formed. The onset of mode penetration approximately takes place when the perpendicular electron velocity at the rational surface has dropped by a factor of two consistent with analytical predictions.

A set of experiments to study the impact of the perturbation amplitude onto mode seeding was performed. The results of these experiments are shown in Fig.~\ref{delay_exp}. The signals of the locked mode detector on the upper part of the plot are shifted in order to match the time of mode penetration across all experiments. The shapes of current amplitude in the RMP coils in the lower part of the Figure are shifted accordingly. The time delay of mode penetration with respect to the RMP ramp-up time increases for lower coil currents, i.e., for lower perturbation amplitudes. Ultimately, when the RMP current becomes too small, no penetration is observed at all. Thus, a threshold in the coil currents is observed for mode penetration (between $100$ and $200\,A$ in this case), below which the $j\times B$ torque does not reduce the perpendicular electron velocity strong enough to reach the transition point.

% ************************************
\section{\label{simresults}Non-linear simulations of mode penetration in ASDEX Upgrade}
% ************************************

In the following, we present simulation results for mode penetration by externally applied resonant magnetic perturbations. This section is organized as follows. In Section~\ref{JOREK}, the non-linear MHD code JOREK used for the simulations is briefly described and the simulation setup is explained in Section~\ref{setup}. An overview of all simulations performed is given in Section~\ref{overview}.

Section~\ref{typical} shows results for a simulation of an ASDEX Upgrade like plasma. All three phases of mode penetration can be seen in this simulation consistent with experiments and analytical theory. Also the evolution of toroidal and poloidal rotation is reproduced qualitatively. Section~\ref{scans} shows results of parameter scans in mode rotation and the coil currents revealing thresholds for mode penetration in both parameters. A hysteresis in island size and plasma rotation between ramp-up and ramp-down of the magnetic perturbation is observed in Section~\ref{hysteresis} consistent with analytical predictions. Finally, Section~\ref{kinktear} investigates the evolution of the kink and tearing responses during mode penetration.

% ************************************
\subsection{\label{JOREK}Physics model and JOREK non-linear MHD code}
% ************************************

Our simulations are performed with the 3D non-linear MHD code JOREK\cite{Huysmans2007} which is routinely applied to a variety of ELM and disruption related questions in tokamak X-point plasmas and has already been used for penetration studies of external magnetic perturbations\cite{NardonNF2010,OrainPoP2013,Becoulet2014}. The code uses 2D bi-cubic B\'{e}zier finite elements in the poloidal plane, and a Fourier expansion in the toroidal direction\cite{Czarny2008}. The physics model used for our investigations is a resistive reduced-MHD\cite{Strauss1997,Franck2015} model with extensions~\cite{Orain2013} for two-fluid effects, realistic neoclassical poloidal rotation, and realistic toroidal background rotation. The main reduced MHD assumption is that the magnetic field is expressed as $\mathbf{B}=F_0/R\;\mathbf{e}_\phi + R^{-1}\;\nabla\Psi\times\mathbf{e}_\phi$ where $F_0=R_0 B_{\phi0}$ is constant in time and space, $R_0$ is the major radius, $\Psi$ is the poloidal flux, $B_{\phi0}$ is toroidal magnetic field amplitude at the magnetic axis, and $\phi$ is the toroidal coordinate.
The basic set of equations solved implicitly in the code includes the continuity equation, the parallel and perpendicular components of the momentum equation, the energy conservation equation, Ohm's law and definition equations for the current and the vorticity. Two-fluid diamagnetic effects are included in the system via the diamagnetic velocity $\mathbf{V}^\ast_s = -\nabla P_s \times \mathbf{B}/(\rho e_s B^2/m_i)$ of each species $s$ - electrons and ions. Here $P_s$  is the pressure of the species $s$, $\rho = m_i n$ is the mass density of the plasma, since electrons are much lighter than ions, the electron contribution is neglected, $n=n_i=n_e$ is the particle density, assuming quasineutrality condition and singly charged ions. $e_s=\pm e$ denotes the electric charge of each species, and $m_i$ is the ion mass. The fluid velocity is the sum of the $\mathbf{E} \times \mathbf{B}$ drift velocity $\mathbf{V}_E = \mathbf{E} \times \mathbf{B}/B^2$, the parallel and ion diamagnetic velocities
\begin{equation}
	\mathbf{V} \approx \mathbf{V}_i = \mathbf{V}_{\|,i} + \mathbf{V}_E + \mathbf{V}^\ast_i
\end{equation}
Neoclassical effects are considered in the momentum equation, where the pressure tensor is given by $\bar{P}=\bar{I}P + \bar{\Pi}_{i,neo} + \bar{\Pi}_{i,gv}$. After gyroviscous cancellation~\cite{Hazeltine_PoF1985} and adopting the expression for the divergence of the neoclassical tensor derived by Gianakon et al.~\cite{Gianakon}
\begin{equation}
    \nabla \cdot \bar{\Pi}_{i,neo}=\rho \mu_{i,neo} \frac{B^2}{B^2_{\theta}}(V_{\theta}-V_{\theta,neo})\mathbf{e_\theta}
\end{equation}
with $\mu_{i,neo}$ being the neoclassical friction and $V_{\theta,neo}=-\kappa_i\nabla T_i \times \mathbf{B}/eB^2 \cdot \mathbf{e}_\theta$, where $\kappa_i$ is the neoclassical heat diffusivity, the final set of model equations reads:
\begin{equation}
	\frac{\partial \rho} {\partial_t} = - \nabla \cdot \Big(\rho \mathbf{V}\Big) + \nabla \cdot \Big(D_{\bot } \nabla_{\bot } \rho \Big) + S_\rho ,
	\label{eqnsystem1}
\end{equation}
\begin{multline}
	\rho \frac{\partial V_{\|,i}}{\partial t} = \mathbf{b} \cdot \Bigg[-\rho\Big( \big(\mathbf V_{\|,i} + \mathbf{V}_E\big)\cdot \nabla\Big)\mathbf{V} - \nabla P \\ - \nabla \cdot \bar{\Pi}_{i,neo} \Bigg] + \mu_\|\Delta \Big(V_{\|,i}-V_{\|,\text{NBI}}\Big),
	\label{eqnsystem2}
\end{multline}
\begin{multline}
%	\begin{split}
	\mathbf{e}_\phi \cdot \nabla \times \Bigg[\rho \frac{\partial \mathbf{V}_E}{\partial t} = - \rho\Big(\mathbf{V \cdot \nabla}\Big)\mathbf{V}_E \\ + \mathbf{J} \times \mathbf{B} - \nabla P - \nabla \cdot \bar{\Pi}_{i,neo} + \mu_\bot\Delta V \Bigg],	
%	\end{split}
\end{multline}
\begin{multline}
	\frac{\partial(\rho T)}{\partial t} = -\rho \mathbf{V_E} \cdot \nabla T - (\gamma - 1)P\nabla \cdot \mathbf{V_E} \\ + \nabla \cdot \Big(\kappa_\|\nabla_\| T + \kappa_\bot\nabla_\bot T\Big) + S_T,
\end{multline}
\begin{equation}
	\frac{1}{R^2}\frac{\partial \psi}{\partial t} = -\mathbf{B} \cdot \nabla_\| u + \frac{\tau_{IC}}{\rho}\mathbf{B}\cdot \nabla_\|P + \frac{\eta}{R^2}\Big(J-J_0\Big),
\end{equation}
with the parallel gradient defined as following
\begin{multline}
	\nabla_\| \alpha = \mathbf{b}(\mathbf{b} \cdot \nabla \alpha)
	= \frac{\mathbf{b}}{B}\left(\frac{F_0}{R^2}\partial_\phi \alpha + \nabla \phi \cdot \nabla \alpha \times \nabla \psi \right)
\end{multline}
In this system of equations, $T$ denotes the temperature (assuming same temperature for ions and electrons), $J$ is the toroidal current, $u$ is the electrostatic potential, $D_\bot$ is the particle diffusion coefficient, $\mu_\|$ and $\mu_\bot$ are
the parallel and perpendicular viscosity coefficients, $\gamma = 5/3$ is the adiabatic index, $\kappa_\bot$ and $\kappa_\|$ are the perpendicular and parallel heat diffusivities, $\eta$ is the
resistivity, $\mathbf{b}=\mathbf{B}/B$ and $B=|\mathbf{B}|$ is the unit vector along the magnetic field, and $\mathbf{e}_\theta=\mathbf{b}\times\mathbf{e}_\phi$ is the poloidal unit vector. In the simulations, both the plasma resistivity and the viscosity evolve in time according to Spitzer-like $(T/T_0)^{3/2}$ dependence and the parallel heat diffusivity evolves as $(T/T_0)^{5/2}$, where $T_0$ is the initial temperature in the plasma centre.

$S_\rho$ and $S_T$ are sources of particles and heat, respectively. 
The terms $V_{\|,\text{NBI}}$ and $J_0$ drive the parallel rotation and the current density in the absence of an island towards the initial profiles by compensating the decay due to parallel viscosity \dmytro{and resistivity respectively}. The radial profile of heat and particle sources is assumed to be Gaussian. Source and diffusion profiles are adjusted to keep the density and temperature profiles close to the initial profiles in the absence of an instability\matthias{, such that the steady state values and gradients of density and temperature remain close to the initial values in the region of interest, i.e., in the vicinity of the $q=2$ resonant surface.}
\matthias{Note, that the employed gyroviscous cancellation breaks the strict conservation of energy. Linear benchmarks have shown that growth rates are correct with our present model. Plans exist to implement a fully conservative model without this cancellation, however this is left for future work. The Ohmic heating term is not accounted for in our simulations, since we take an artificially increased resistivity (see next Section), which would lead to an unrealistic source of thermal energy. Instead the thermal energy is modelled via a Gaussian source.}

The perturbation induced by RMP coils is modeled as Dirichlet boundary condition for the poloidal magnetic flux. A pure $n=1$ perturbation is applied. The vacuum RMP spectrum is calculated at the boundary of the JOREK computational domain with an external program and applied as Dirichlet boundary condition. RMPs are progressively switched on in time: the amplitude of the perturbation is gradually increased within a typical timescale $t \sim 1000\tau_A$, where $\tau_A=\sqrt{\mu_0 \rho_0}$ approximates the Alfven time scale, $\mu_0$ is the vacuum permeability, and $\rho_0$ the mass density in the plasma center. For the plasma parameters of our equilbrium, $\tau_A=0.6\mu s$. This way, the magnetic perturbation gradually penetrates into the plasma, which self-consistently adapts in the process. \dmytro{The choice of RMP timescale was taken for numerical reasons and is not consistent with the experimental ordering with typical visco-resistive and toroidal rotation times, however this does not affect our physical results as we have checked by varying the ramp-up function.}

% ************************************
\subsection{\label{setup}Simulation setup}
% ************************************

\begin{figure}[htbp]
    \includegraphics[height=9em]{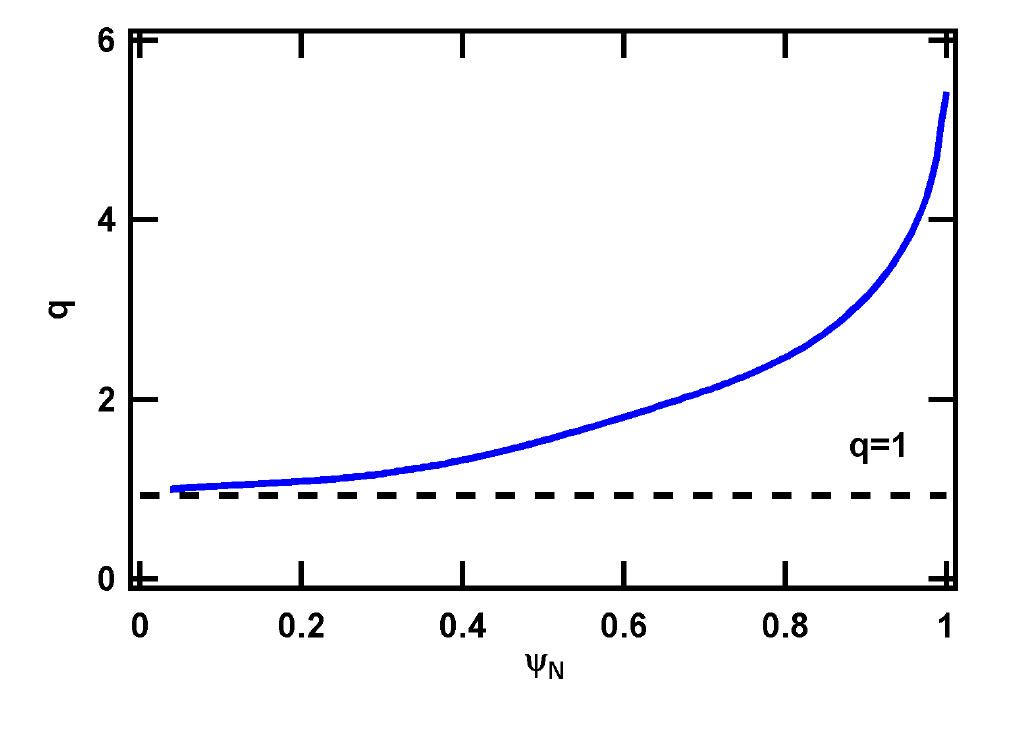}
	\includegraphics[height=9em]{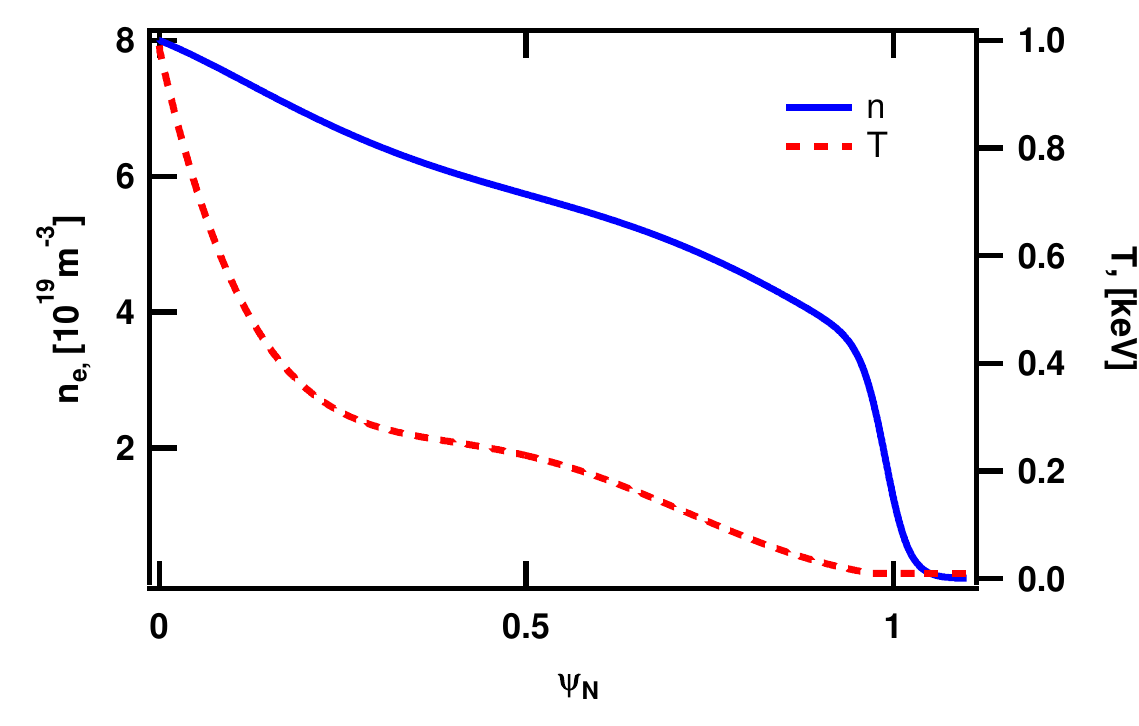}
	\caption{Profiles of the safety factor q (top) and the density and temperature (bottom ).  All simulation input is based on the CLISTE equilibrium reconstruction for ASDEX Upgrade L-Mode discharge \#30734 at 1.2 seconds.}
	\label{input-profiles}
\end{figure}

\begin{figure}[htbp]
    \centering
	\includegraphics[width=0.8\linewidth]{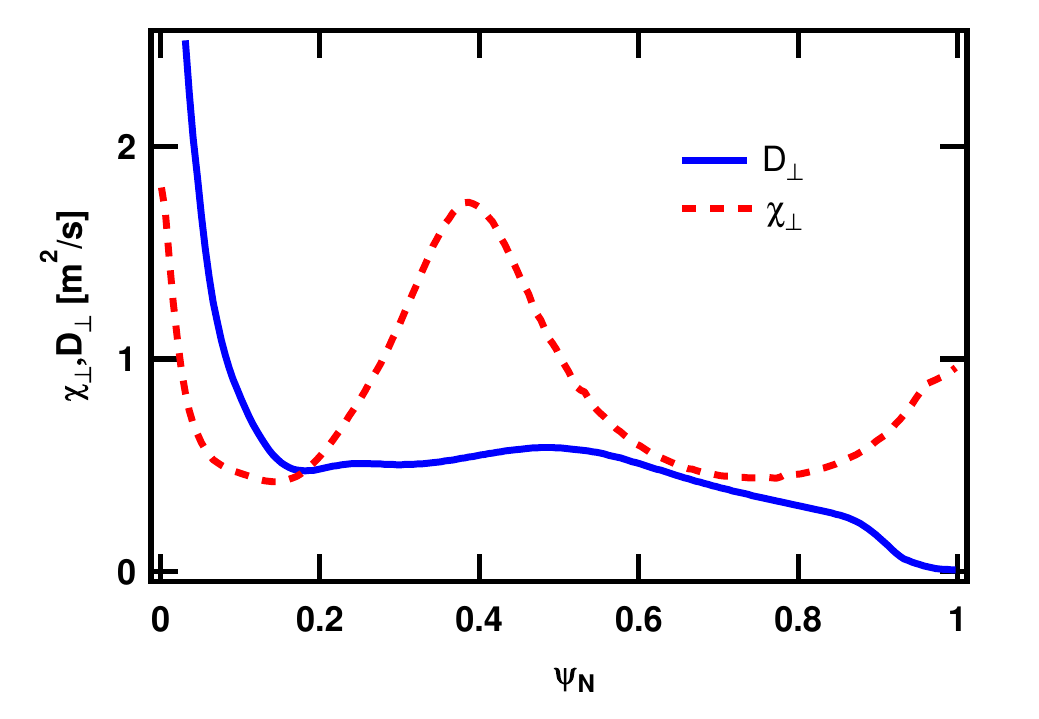}	
	\caption{Profiles of the perpendicular heat and particle diffusivities used in  the simulations. Note that $\chi_\bot$ is connected to $\kappa_\bot$ used in the model equations by $\chi_\bot=\kappa_\bot/\rho$ where $\rho$ denotes the mass density.}
	\label{diff-profiles}
\end{figure}

In our simulations, JOREK is initialized for an ASDEX-Upgrade like plasma: major radius $R\approx 1.65\,m$, minor radius $a\approx 0.5\,m$, toroidal field strength on axis $B_t=1.9\,T$, plasma current $I_p=1\,MA$, edge safety factor $q_{95}=3.8$. The full X-point plasma including the scrape-off layer up to simplified divertor targets are included in the simulation domain. The central electron density is $n_{e,0}=8\cdot10^{19}\,m^{-3}$, the central temperature is $T_{e,0}=1keV$. The perpendicular heat diffusion coefficient is in the range $\chi_\bot \sim 0.5\,m^2/s$. The parallel heat diffusion coefficient is proportional to $T_e^{5/2}$ with the value $\chi_\| \sim 3.5\cdot10^8\,m^2/s$ at the plasma center chosen to be about one order of magnitude lower than Spitzer-H\"arm predictions~\cite{Spitzer1953} since typically the heat flux limit~\cite{Malone1975} reduces the parallel conductivity in the experiment~\cite{Hoelzl2009}. 
RMP currents are chosen around $I_{RMP}\sim 1kA$ and scanned in a few simulations. The perpendicular electron background velocity is modified by changing the toroidal rotation velocity. Input profiles are given in Fig.~\ref{input-profiles}. Input heat and particle diffusion profiles are shown in Fig.~\ref{diff-profiles}. The Lundquist number $S=\frac{\mu_0 a v_\text{Alfv\'en}}{\eta}\approx1\cdot10^7$ in the simulations is close to experimental conditions. \matthias{The magnetic Prandtl number is chosen to be constant across the whole simulation domain such that viscosity is close to the collisional value estimated from the experimental data $P_{rm}=\nu/\eta=(\mu/\rho_0)/\eta=10$.}

The most important effect missing in our simulations is the neoclassical toroidal viscosity, which would lead to a faster penetration of the magnetic perturbation and a penetration already at lower RMP coil currents or higher initial plasma rotation frequencies. We leave the implementation and study of this effect for future studies. A consistent evolution of the bootstrap current, which is available in the JOREK code for further studies, is also neglected since we are interested in particular in the seeding and penetration phases. In addition, the hysteresis effect investigated in Section~\ref{hysteresis} can be studied this way independently of the the bootstrap current term simplifying the interpretation of the results.

\begin{figure}[htbp]
	\includegraphics[height=20em]{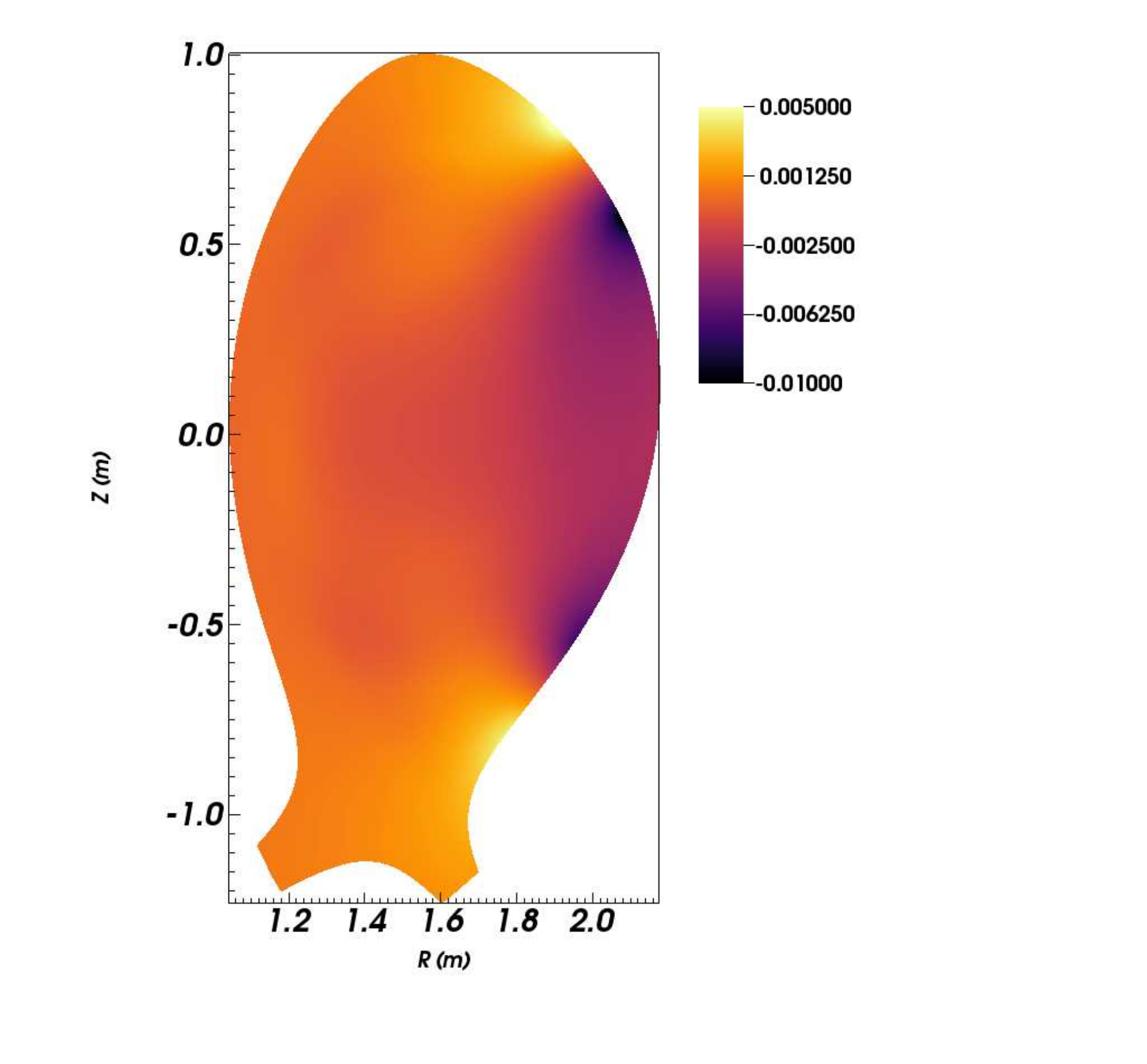}
		\caption{The vacuum poloidal flux is prescribed at the boundary of the computational domain for simulating the RMP coils.}
\end{figure}

% ************************************
\subsection{\label{overview}Overview of the simulations performed}
% ************************************

\begin{table}[htbp]
\caption{\label{tab:table1} Simulations performed for the scan in RMP current amplitude}%
\begin{ruledtabular}
	\begin{tabular}{c|cccc}
	    $V_{tor}$ & & \textrm{$I_{RMP}$}\\
	    \colrule
		\textrm{$2.5\,km/s$}& \textrm{$1.9\,kA$}& \textrm{$2\,kA$}\\
		\textrm{$6.2\,km/s$}& \textrm{$2\,kA$}& \textrm{$2.2\,kA$}& \textrm{$2.4\,kA$}\\
	\end{tabular}
\end{ruledtabular}
\end{table}

\begin{table}[htbp]
\caption{\label{tab:table2} Simulations performed for the scan in the plasma rotation}%
\begin{ruledtabular}
	\begin{tabular}{c|ccc}
    	$I_{RMP}$ & & \textrm{$V_{tor},\,[km/s]$}\\
	    \colrule
		\textrm{$2\,kA$}& \textrm{$2.5$}& \textrm{$4.4$}& \textrm{$6.2$}\\
	\end{tabular}
\end{ruledtabular}
\end{table}

\begin{table}[htbp]
\caption{\label{tab:table4} Correspondence of the values of perpendicular electron velocity $V_{\bot,e}$ at the $q=2$ resonant surface to the values of toroidal background rotation $V_{tor}$ used in the parameter scan}%
\begin{ruledtabular}
	\begin{tabular}{c|ccc}
		\textrm{$V_{tor}\,[km/s]$}&
		\textrm{$2.5$}&
		\textrm{$4.4$}&
		\textrm{$6.2$}\\
		\textrm{$V_{\bot,e}\,[km/s]$}&
		\textrm{$2.7$}&
		\textrm{$3$}&
		\textrm{$3.3$}\\
	\end{tabular}
\end{ruledtabular}
\end{table}

In the following, we give a brief overview of the simulations performed for this paper and refer to the respective sections, in which the results are discussed in detail. 

Simulations at fully realistic Lundquist number for ASDEX Upgrade L-Mode experiments have been performed and reflect the qualitative change of the toroidal and electron perpendicular rotation profiles, however full penetration was not obtained (very likely due to the missing NTV effects).
Thus, all simulations shown in the following are performed with values reduced by about a factor three to $S=1 \cdot 10^7$.

The typical process of mode seeding is shown and analyzed in~\ref{typical}. The initial toroidal velocity used in the simulations is $V_{tor}=2.8\,km/s$ and RMP current $I_{RMP} = 2\,kA$. The thresholds for mode seeding in perturbation amplitude and rotation are studied in Section~\ref{scans} by means of the parameter scans shown in Table~\ref{tab:table1} and Table~\ref{tab:table2}. The toroidal rotation was used as proxy to modify the perpendicular electron velocity as shown in Table~\ref{tab:table4}.

Finally, simulations to study the hysteresis behaviour between ramp-up and ramp-down of the RMP coil currents are shown in section~\ref{hysteresis}. These simulations were performed with $V_{tor}=6.2\,km/s$ and the RMP current amplitudes shown in Table~\ref{tab:table3}.

\begin{table}[htbp]
\caption{\label{tab:table3} Simulations performed for the hysteresis studies}%
\begin{ruledtabular}
	\begin{tabular}{c|ccccccc}
	    & & & \textrm{$I_{RMP}$}\\
	    \colrule
		\textrm{Ramp-up}& \textrm{$2\,kA$}& \textrm{$2.2\,kA$}& \textrm{$2.4\,kA$}&\\
		\textrm{Ramp-down}& \textrm{$2.2\,kA$}& \textrm{$2\,kA$}& \textrm{$1.9\,kA$}& \textrm{$1.75\,kA$}& \textrm{$1.5\,kA$}& \textrm{$1\,kA$}&\\
	\end{tabular}
\end{ruledtabular}
\end{table}

% ************************************
\subsection{\label{typical}Typical simulation of mode penetration into an ASDEX Upgrade L-Mode plasma}
% ****************************

\begin{figure}
	\centering
	\includegraphics[width=0.8\linewidth]{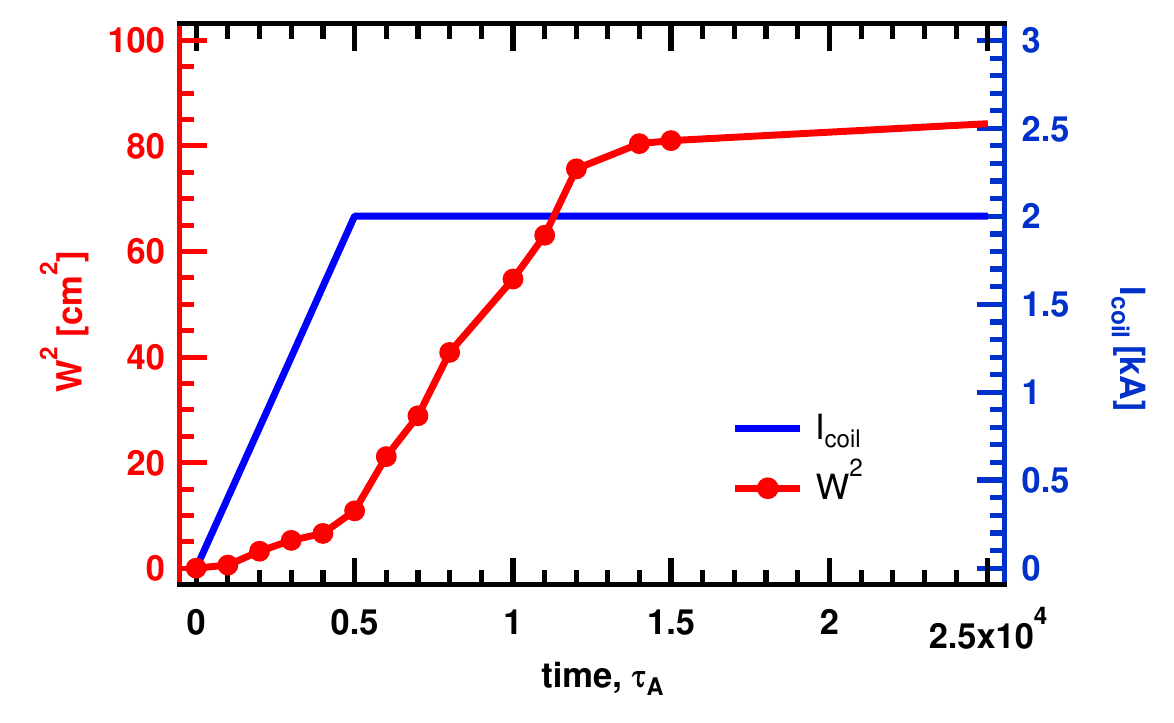}
	\caption{Square of the island size~(left axis) and current in the RMP coils~(right axis).}
	\label{W2_t_1e7_V01}
\end{figure}

\begin{figure}[htbp]
	\centering
	\includegraphics[width=0.8\linewidth]{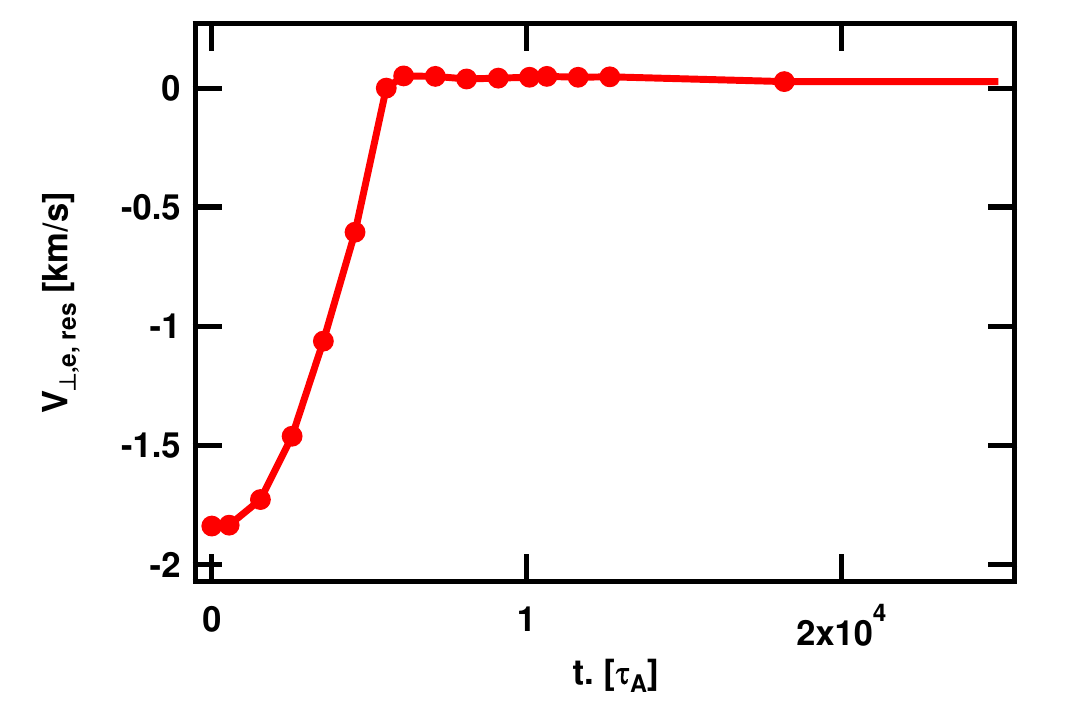}	
	\caption{Time evolution of the perpendicular electron velocity profiles from JOREK simulations with $S=10^7$ at the $q=2$ resonant surface.}
	\label{Vpe_res}
\end{figure}

This section shows a typical simulation of mode penetration performed with $V_{tor}=2.8\,km/s$ and $I_{RMP}=2\,kA$ at $S=10^7$.
Like in the experimental observations shown in Section~\ref{exp}, three phases are observed. First, the plasma exhibits a weak response to the applied perturbation. Once a specific threshold is reached (see Figure~\ref{om_Wvac_om0}), mode growth accelerates until the final mode saturation at low perpendicular electron velocity is obtained.
Time traces for the evolution of the island size and the perpendicular electron velocity at the rational surface are shown in Figures~\ref{W2_t_1e7_V01} and Figures~\ref{Vpe_res}, respectively.

\begin{figure}[htbp]
	\centering
	\includegraphics[width=0.8\linewidth]{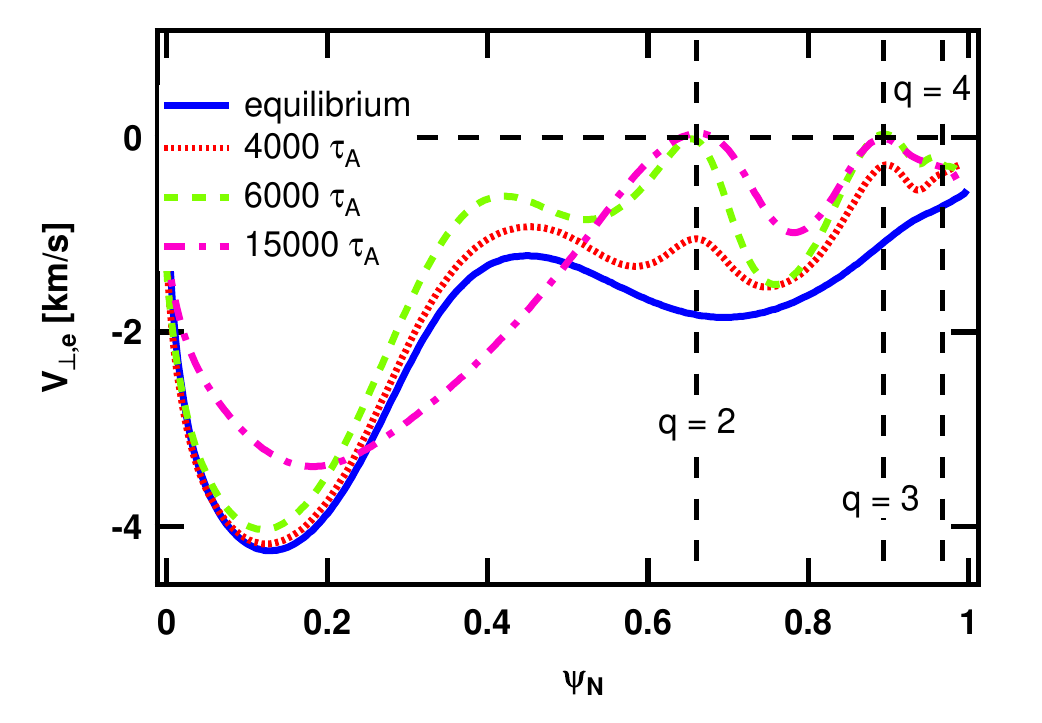}	
	\caption{Perpendicular electron velocity profiles from JOREK simulations with $S=10^7$}
	\label{Vpe_sim_S1e7}
\end{figure}

In the beginning, the island size is small and the perpendicular electron velocity at the rational surface is around $\sim 1.8\,km/s$. The electron velocity is given by $\mathbf{V}_e=\mathbf{E}\times\mathbf{B}/B^2+\nabla p\times\mathbf{B}/(2 n_e e B^2)+\mathbf{V}_{||,e}$, where the factor $2$ is a result of the assumption of having the same electron and ion temperatures in the model employed here (a model treating both temperatures differently is available as well).
The rotation velocity has dropped by a factor of two after approximately $t=4100\tau_A$ and an accelerated growth of the island (mode penetration) sets in at around $t=4900\tau_A$. The electron velocity comes to rest at the rational surface around $t=6100\tau_A$ and the saturation of the island starts around $t=12000\tau_A$. Figure~\ref{W2_t_1e7_V01} shows the time evolution for the square of the island width which is proportional to the current perturbation at the resonant surface and, consequently, approximately proportional to the expected signals in magnetic pick up coils. The Figure also contains the prescribed evolution of the RMP coil currents. Note, that it is coincidence that saturation of the prescribed coil currents and onset of mode penetration take place approximately at the same time for this particular case (as proven by other simulations in our scans). For the perpendicular electron velocity at the rational surface shown in Figure~\ref{Vpe_res}, also radial profiles at several points in time are provided in Figure~\ref{Vpe_sim_S1e7}. Profiles are shown in the beginning of the simulation, before mode penetration sets in, during mode penetration, and in the saturated island state.

\begin{figure}[htbp]
	\centering			\includegraphics[width=0.8\linewidth]{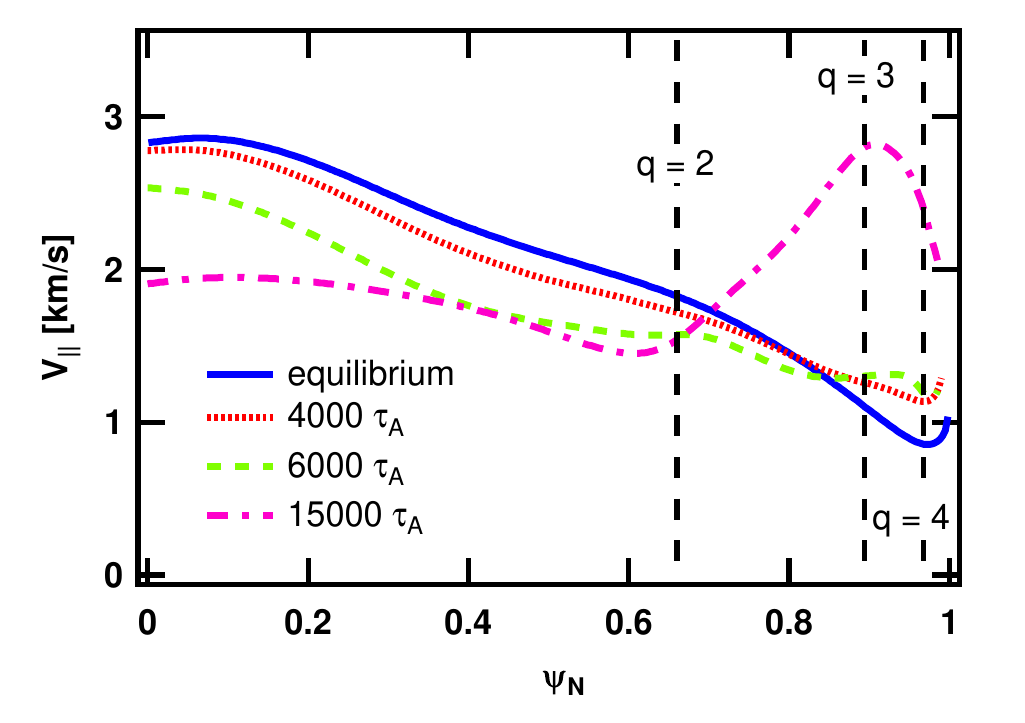}
	\caption{Toroidal rotation velocity profiles from JOREK simulations with $S=10^7$}
	\label{Vpar_sim_S1e7}
\end{figure}

\begin{figure}
	\centering
	\includegraphics[width=0.8\linewidth]{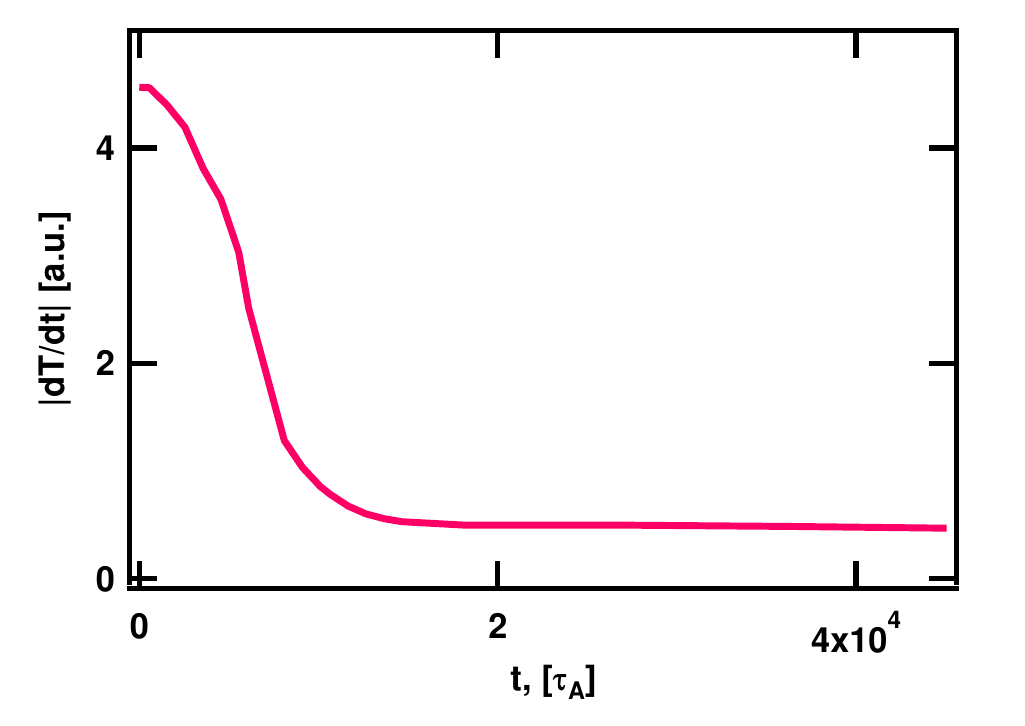}
	\caption{Time evolution of \dmytro{poloidally and toroidally averaged} temperature gradient $dT/d\Psi_N$ at the $q=2$ resonant surface.}
	\label{dTres}
\end{figure}

Initially the non-zero electron perpendicular velocity leads
to the screening of the magnetic perturbation. However, the screening starts to drop due to the loss of the perpendicular component of the toroidal velocity (see Figure~\ref{Vpar_sim_S1e7}) and flatting of the temperature (see Figure~\ref{dTres}) leading to a partial loss of the diamagnetic component. Once the condition $V_{\bot,e}=0$ is approximately satisfied, the transition phase is reached. In this phase, the perturbation propagates without screening forcing magnetic reconnection at the resonant surface. Similar to the experimental observations, the core toroidal velocity shown in Fig.~\ref{Vpar_sim_S1e7} decreases. %At the same time the edge toroidal velocity increases. 

\begin{figure}[htbp]
    \centering
    \includegraphics[width=0.8\linewidth]{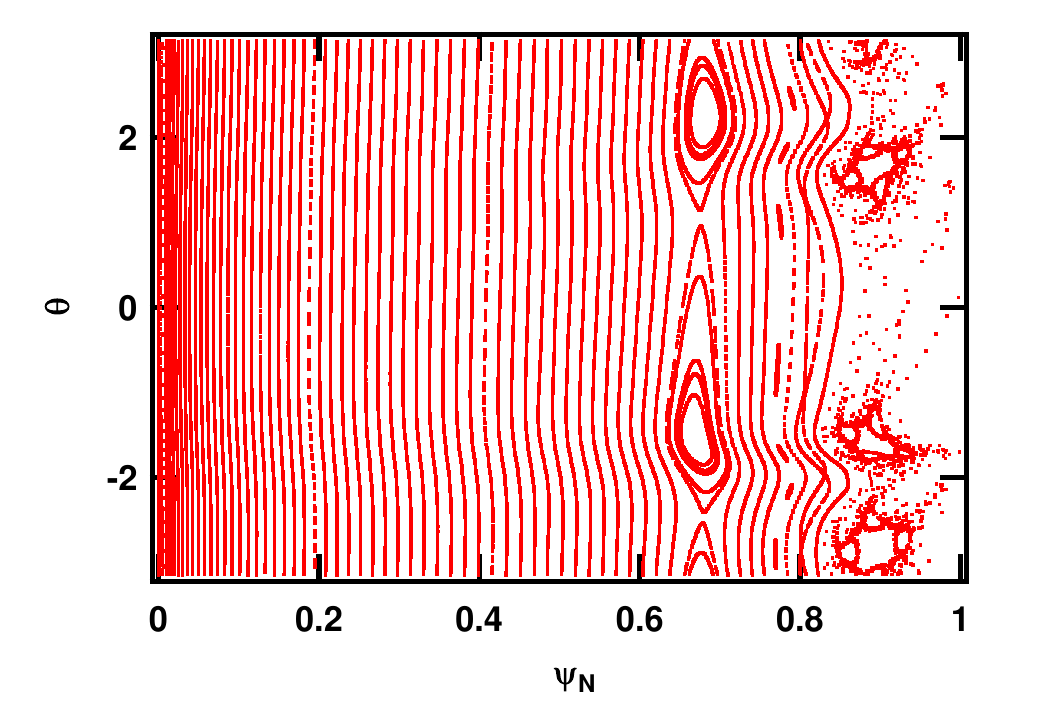}	
	\includegraphics[width=0.8\linewidth]{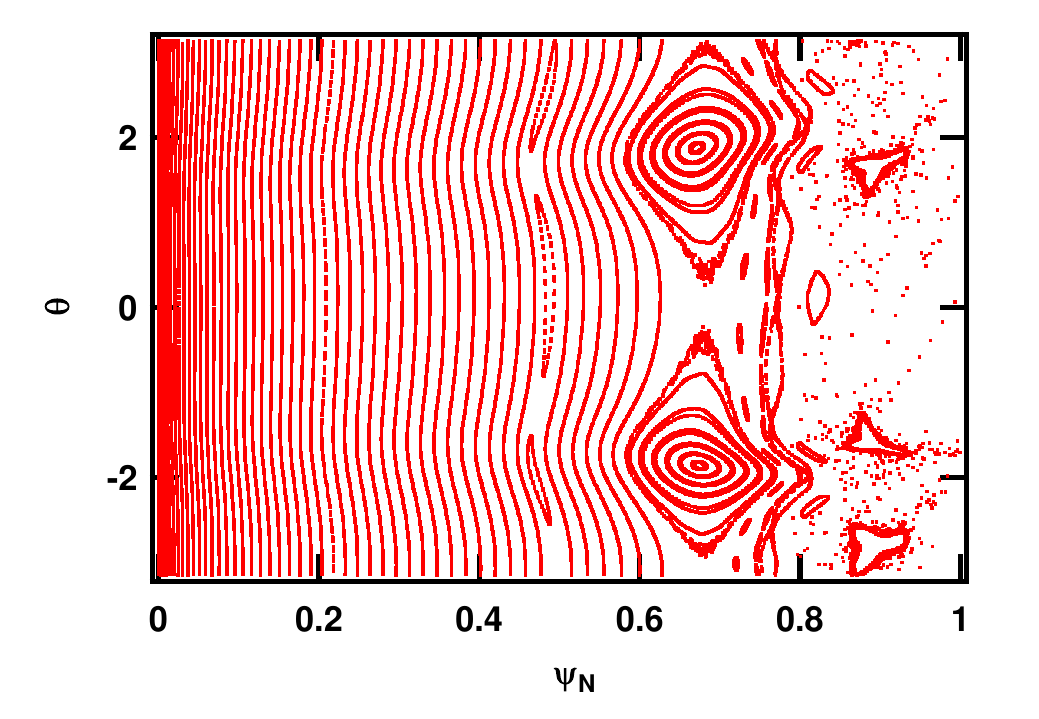}	
	\caption{Poincar\'{e} plots of the magnetic topology in ($\psi_N,\theta$) coordinates before mode penetration $t=4000\tau_A$~(upper plot) and in fully formed island phase $t=15000\tau_A$~(lower plot). The width of the $q=2$ island on the upper plot $W=3.8\,cm$ and on the lower one $W=9.2\,cm$. \matthias{The 3/1 and 4/1 islands penetrate faster than the 2/1 island since the applied spectrum contains significant 3/1 and 2/1 components and the electron velocity is slow at the respective surfaces such that shielding is less effective.}}
	\label{poinc}
\end{figure}

Figure~\ref{poinc} finally shows Poincare plots of the magnetic topology before mode penetration sets in and in the saturated island state. A considerable stochastisation of the plasma edge, and a large $2/1$ magnetic island can be seen.

% ************************************
\subsection{\label{scans}Mode penetration thresholds in coil current and rotation velocity}
% ************************************

Scans in both the plasma rotation velocity and the perturbation amplitude (coil currents) were carried out. The electron perpendicular velocity is modified in our scan by changing the toroidal velocity, while keeping the 
$\mathbf{E}\times\mathbf{B}$ and diamagnetic drift effects unchanged. Table~\ref{tab:table4} shows how the perpendicular electron velocity is affected by our choice of the toroidal rotation velocity.

\begin{figure}[htbp]
	\centering
        \includegraphics[width=0.8\linewidth]{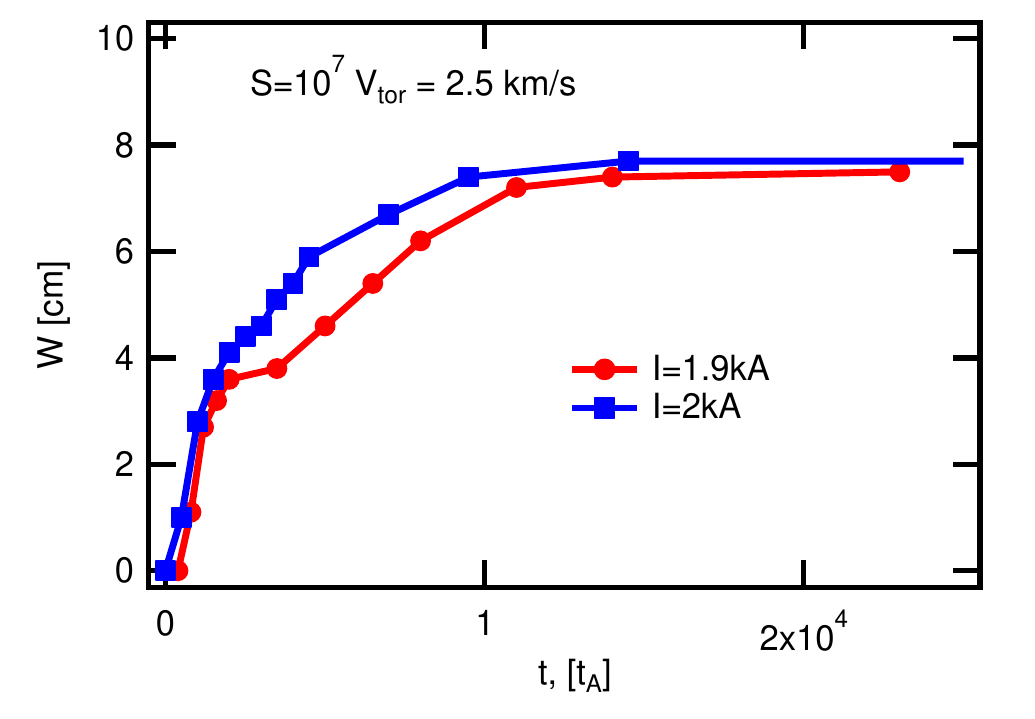}	
		\caption{Island size evolution for different values of the current in RMP coils at $V_{tor} = 2.5\,km/s$}
		\label{WvsT_Iscan_V01}
\end{figure}

\begin{figure}[htbp]
		\centering
    	\includegraphics[width=0.8\linewidth]{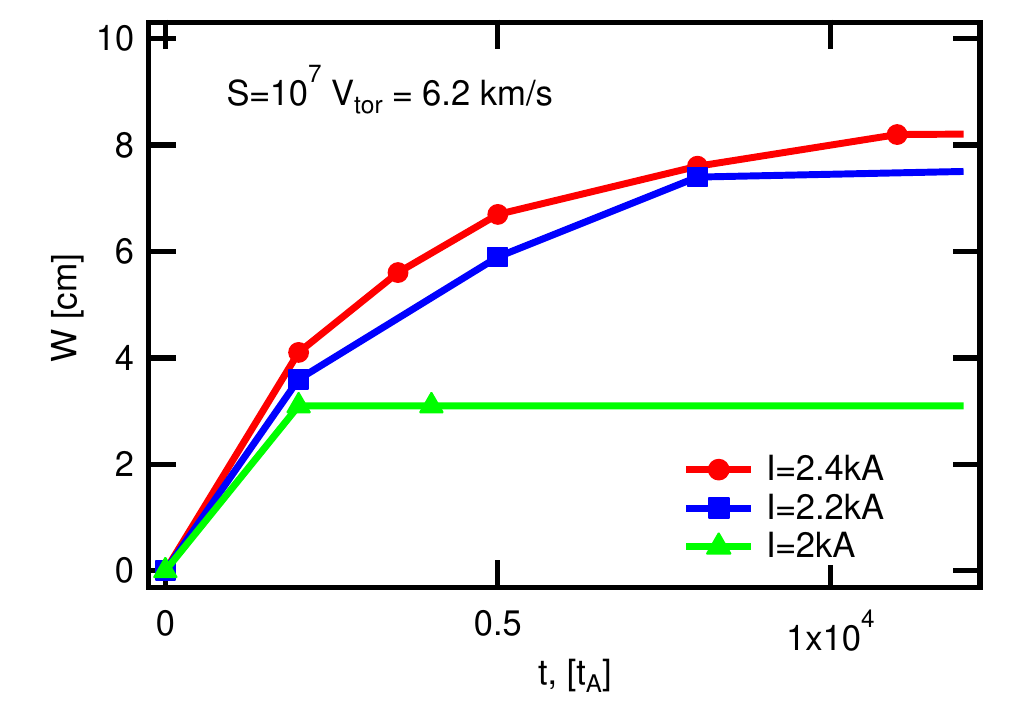}	
		\caption{Island size evolution for different values of the current in RMP coils $V_{tor} = 6.2\,km/s$}
		\label{WvsT_Iscan_V03}
\end{figure}

Scans in the perturbation amplitude for the values of toroidal velocity $V_{tor}=2.5\,km/s$ and $V_{tor}=6.2\,km/s$ are shown in Fig.~\ref{WvsT_Iscan_V01} and Fig.~\ref{WvsT_Iscan_V03} respectively. As the current in the RMP coils decreases, mode penetration slows down consistently with experimental observations (see Section~\ref{exp}). Ultimately, if the perturbation is not strong enough, the electromagnetic torque cannot reduce the plasma rotation sufficiently to enter the penetration phase. The transition from weak response phase to the fully formed island phase was observed experimentally~\cite{Fietz_EPS2015}. Also the time, required for the mode penetration, increased with decrease of the RMP current.  \dmytro{ The threshold for the transition to the penetrated state can only be compared qualitatively to the experimental observations due to limitations of our model. In particular, we believe that the lack of Neoclassical Toroidal Viscosity (NTV) in the simulations is our biggest limitation as it would provide a localized decrease in the rotation of the mode.}

\begin{figure}[hbtp]
	\centering
	\includegraphics[width=0.8\linewidth]{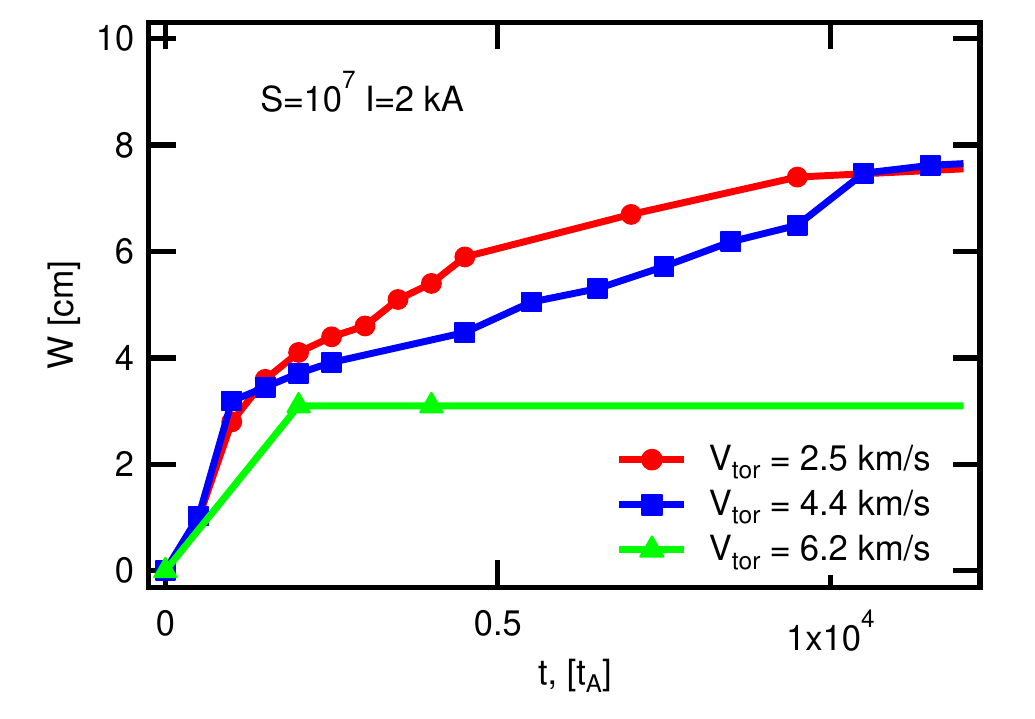}
	\caption{Island size evolution for different toroidal velocities}
	\label{WvsT_Vscan}
\end{figure}

The scan in the plasma rotation shows a delay in the mode penetration as the initial toroidal velocity increases as seen in Figure~\ref{WvsT_Vscan}. This is similar to moving from a lower to an upper curve in Figure~\ref{om_Wvac_om0}. As predicted by the analytical model, if the rotation is strong enough for a given perturbation amplitude, there is no sufficient slow down of the plasma, and the RMP field thus remains partially screened like for the lower curve in Fig.~\ref{WvsT_Vscan}.

It is interesting to point out that, in the case of mode penetration, the final saturated island size is defined solely by the current in RMP coils and is independent from the original plasma velocity. This can be seen in Figure~\ref{WvsT_Vscan}, where the island size follows a universal square-root behaviour with respect to the applied coil currents for the penetrated states.

% ************************************
\subsection{\label{hysteresis}Simulation of the hysteresis in mode penetration between current ramp-up and ramp-down}
% ************************************

\begin{figure}[htbp]
	\centering
    \includegraphics[width=0.8\linewidth]{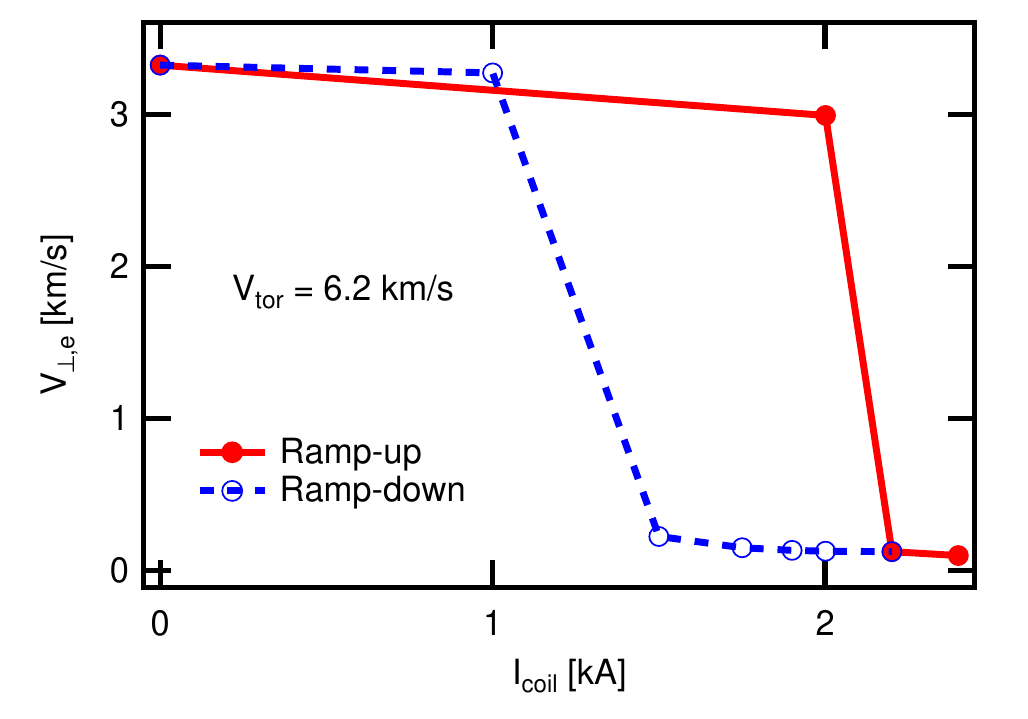}
	\caption{Hysteresis behaviour of the steady state plasma rotation. Initial toroidal rotation velocity $V_{tor}=6.2\,km/s$}
	\label{Vsat_I_hyst}
\end{figure}

\begin{figure}[htbp]
	\centering
    \includegraphics[width=0.8\linewidth]{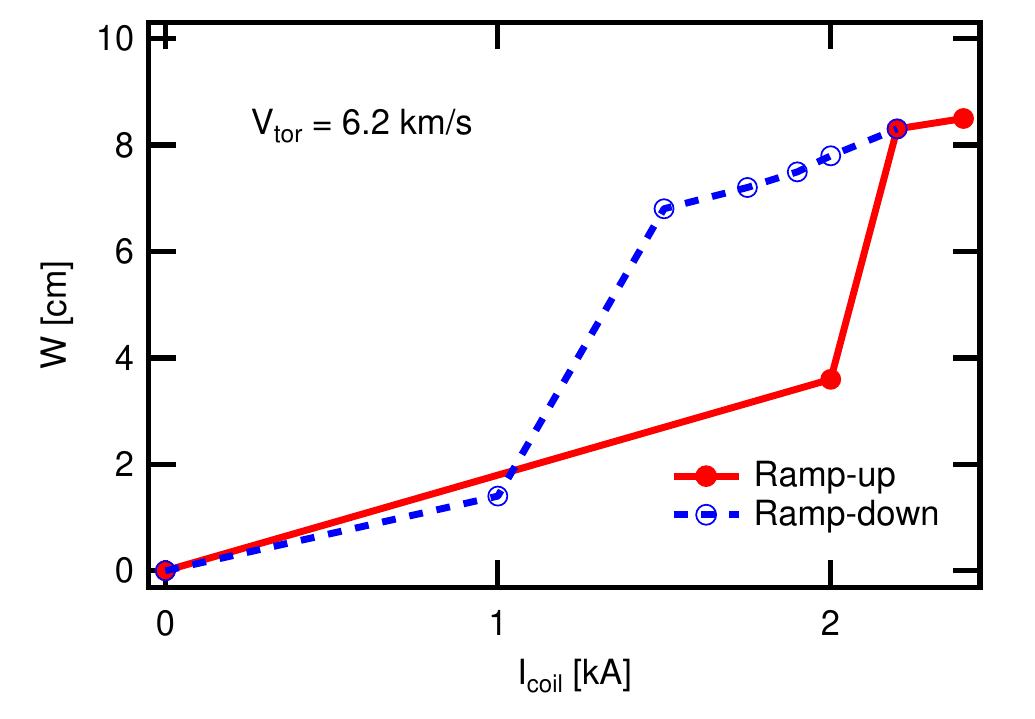}
	\caption{Hysteresis behaviour of the magnetic island size. Initial toroidal rotation velocity $V_{tor}=6.2\,km/s$}
	\label{Wsat_I_hyst}
\end{figure}

An additional set of simulations was performed to study the analytically predicted hysteresis between current ramp-up and ramp-down. As RMP current and associated $\mathbf{j}\times\mathbf{B}$ torque increases, the steady state value for the perpendicular electron velocity at the rational surface gradually decreases. As seen for the solid curve in Figure~\ref{Vsat_I_hyst}, a "jump" is observed to a low rotational state at a coil current of about $2.1 kA$. As seen in Figure~\ref{Wsat_I_hyst}, this corresponds to a "jump" to a large island size, thus to mode penetration.

When the coil current is ramped down now again starting from the steady state solution of the simulation with $2.2 kA$ coil current, the island remains in the penetrated state with low perpendicular electron velocity at the rational surface and large island size significantly longer (dashed line in Figures~\ref{Vsat_I_hyst} and~\ref{Wsat_I_hyst}). The back-transition appears only around a coil current of $1.2 kA$. This implies that a decrease of the perturbation amplitude only slightly below the penetration threshold does not cause a significant decrease of the island size. The small drop of the island size is only given by the square-root dependency of the penetrated island size to the perturbation amplitude. In case of significant pressure gradients, the additional bootstrap current drive can lead to a non-linearly unstable island such that the island remains present even if the external perturbation is switched off entirely again (NTM). The bootstrap current drive is neglected in the present study due to the low pressure gradients in the considered L-Mode plasma.

% ************************************
\subsection{Evolution of \matthias{kink/bending} and tearing responses during mode penetration}\label{kinktear}
% ************************************

\begin{figure}[htbp]
	\centering
	\includegraphics[width=0.8\linewidth]{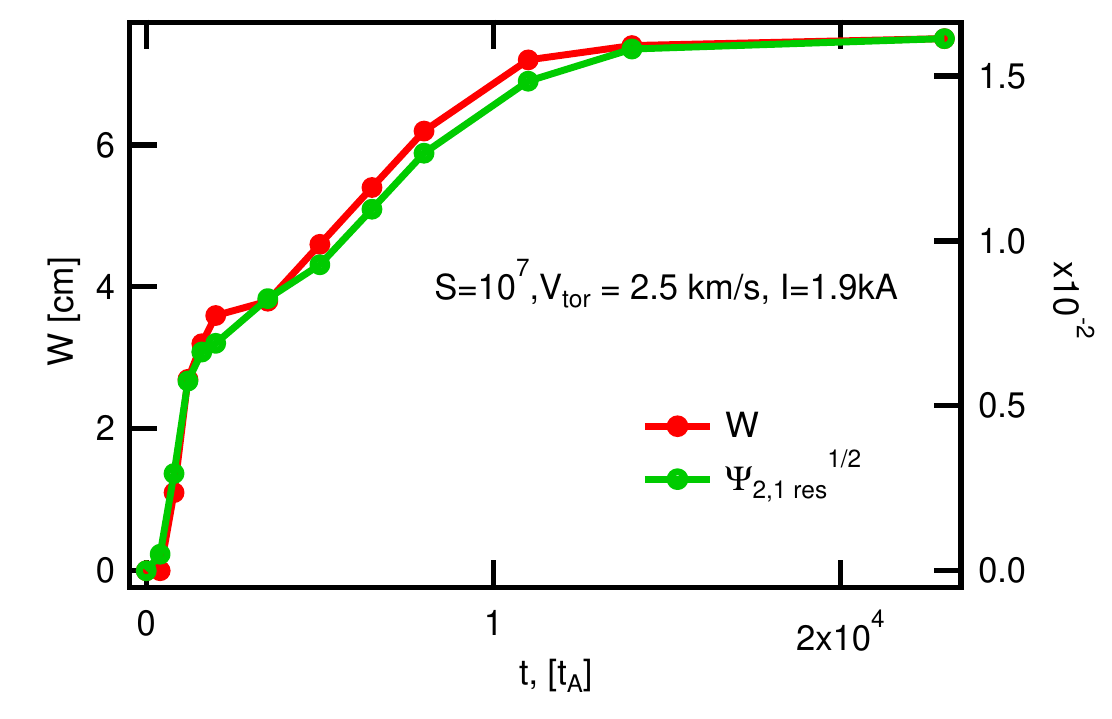}
	\caption{Time evolution of the island size and the $2/1$ component of the magnetic flux}
	\label{WPsi_21_vs_T}
\end{figure}

In this Section, the transition from \matthias{kink/bending}\footnote{the term bending is preferred e.g.~in Ref~\cite{Ferraro2012}) over kinking; we use both terms synonymously since both of them are common in literature} to tearing response is briefly investigated for one of our simulations. This topic was discussed before e.g., in Refs.~\cite{Igochine_PoP2014,Igochine_NF2017,Yu_NF2012}. Figure~\ref{WPsi_21_vs_T} shows the evolution of the island size over time across the mode penetration time. At the same time, the square-root of $\Psi_{2/1}$ at the rational surface is plotted. Both curves show excellent agreement, verifying that the perturbed poloidal magnetic flux at the rational surface is a very good measure for the island size. This, however is not necessarily true for magnetic measurements at coil locations where the decay of the signals has to be taken into account properly and the plasma response between island location and measurement location can additionally alter the signals.

\begin{figure}[htbp]
	\centering
	\includegraphics[width=0.8\linewidth]{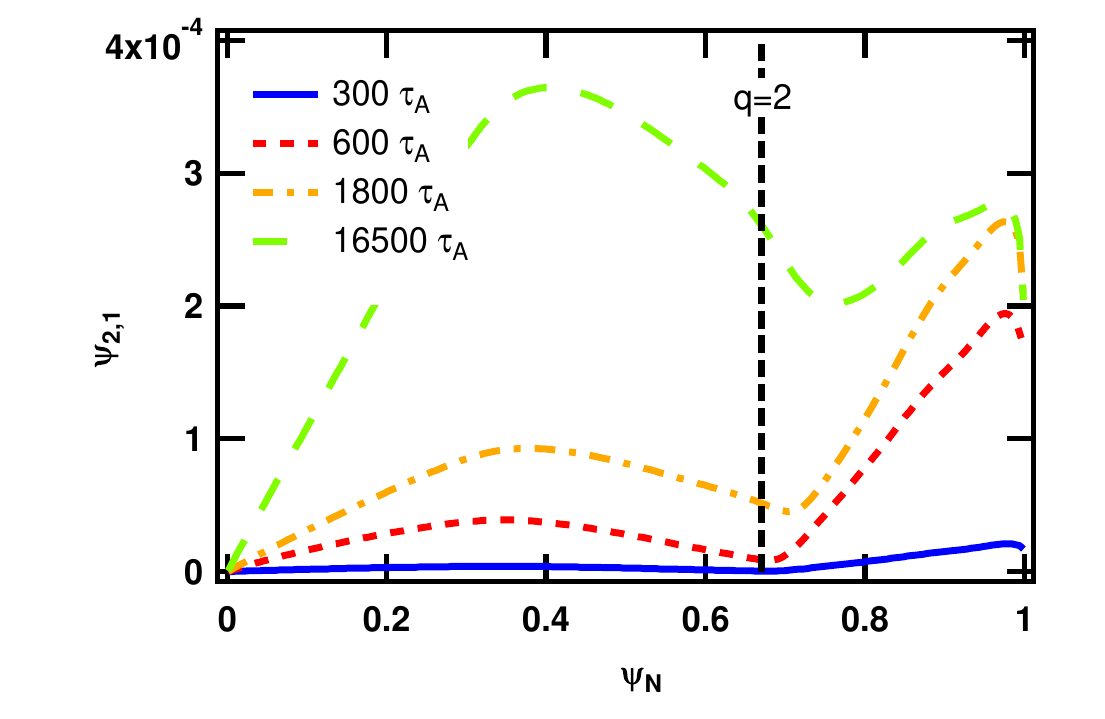}
	\caption{Radial profiles for the absolute value of the $2/1$ component of the poloidal magnetic flux at different time points in the simulation.} %with~(upper) and witout~(lower) mode penetration.}
	\label{Psi_21_vs_T}
\end{figure}

\begin{figure}[htbp]
	\centering
	\includegraphics[width=0.8\linewidth]{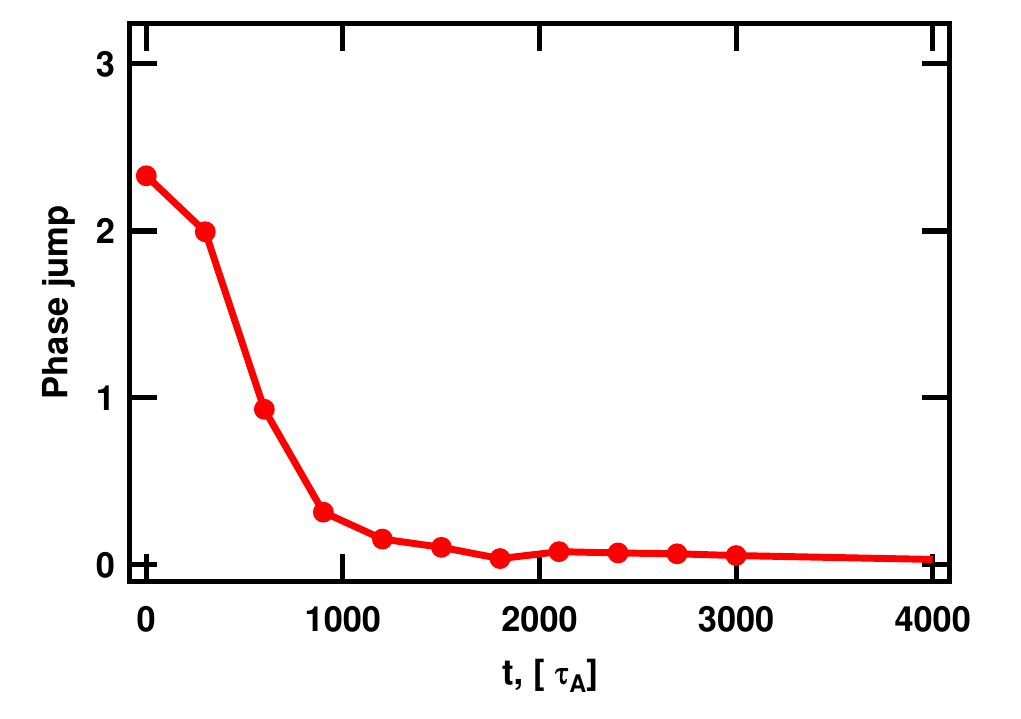}
	\caption{\matthias{Time evolution of the phase jump of the 2/1 component of the poloidal flux across the $q=2$ resonant surface. The difference of the phase of the 2/1 $\Psi$ component between the radial locations $\Psi_N = 0.7$ and $\Psi_N = 0.64$ is plotted. A phase jump close to $\pi$ refers to a dominant kink/bending parity, while a phase jump close to 0 corresponds to a dominant tearing parity.}}
	\label{phase_jump}
\end{figure}

Figure~\ref{Psi_21_vs_T} shows how the absolute value of the $2/1$ magnetic perturbation grows with time. In Figure~\ref{phase_jump} it is depicted how the phase jump of this $2/1$ perturbation across the rational surface is changing with time. \matthias{With phase jump, we refer to the difference of the phase of the $2/1$ poloidal flux between locations outside the rational surface ($\Psi_N = 0.7$) and inside the rational surface ($\Psi_N = 0.64$).} A phase jump close to $\pi$, as it is seen initially, indicates a dominant \matthias{kink/bending} parity of the magnetic perturbation. A phase jump close to zero, like it is approached already after about 1000 Alfv\'en times indicates a dominant tearing parity of the magnetic perturbation. 

%\begin{figure}[htbp]
%	\centering
%	\includegraphics[width=0.8\linewidth]{figures/WPsi_21_n1_vs_T}	
%	\caption{\textit{Time evolution of the island size, the $m=2, n=1$ component and $n=1$ component of the magnetic flux}}
%	\label{WPsi_21_vs_T}
%\end{figure}
%\begin{figure}
%	
%			%\vspace{-2.5cm}
%			\centering
%			\includegraphics[width=0.45\linewidth]{figures/Wsat_I}
%		%	\vspace{-1.cm}		
%			\caption{ \textit{Saturated island size  values of the current in RMP coils}}
%		%	\vspace{-0.5cm}
%			\label{}
%	
%\end{figure}
%
%\begin{figure}[htb]
%%\vspace{-2.cm}
%%	\begin{minipage}[h]{0.83\textwidth}
%		\centering
%%			\vspace{0.5cm}
%			\includegraphics[width=0.83\linewidth]{figures/IcWWmagvsT}	%
%%			\vspace{-0.5cm}		
%			\caption{ \textit{Time evolution of the current in RMP coils, $n=1$ magnetic energy and island size}}
%%			\vspace{-0.5cm}
%				\label{WmWIvst_exp}
%\end{figure}

% ************************************
\section{\label{summary}Conclusions}
% ************************************

Tearing mode seeding by magnetic perturbations has been studied in tokamak X-point geometry both experimentally and numerically and was compared to analytical theory. As a source for the magnetic perturbation, resonant magnetic perturbation coils are used, although our results are fully transferable to other sources like sawtooth crashes~\cite{Loizu_PoP2017}.
The simulations were carried out with the non-linear two-fluid MHD code JOREK. Input parameters close to the experimental ones were chosen, however simulations were performed at slightly reduced Lundquist number and increased coil currents. This was done to compensate for missing NTV effects not accounted for in the simulations.

All three phases of mode penetration observed in the experiment were also obtained in the simulations: "weak" response, a fully formed island state and the transition between these two regimes called penetration. A drop of the core toroidal rotation 
during the penetration is also observed, similar to experiments, however the drop in the simulations is weaker due to the absence of NTV in our simulations. The decay of the electron perpendicular velocity is consistent with experimental observations, meaning that the drop of $V_{\bot,e}$ to 0 corresponds to the mode penetration in both cases. The simulation results were compared to the analytical model for MP penetration derived in cylindrical geometry. Scans in toroidal plasma rotation and perturbation amplitude confirm the analytically predicted thresholds for the fast  transition into low rotation regime. A hysteresis between the RMP current ramp-up and ramp-down was observed like analytically predicted as well. We confirmed a fast formation of the \dmytro{kink/bending} response and a delayed tearing response at the rational surface.

% ************************************
\section*{Acknowlegments}
% ************************************

 This work was carried out under the auspices of the Max-Planck-Princeton Center for Plasma Physics. This work has been carried out within the framework of the EUROfusion Consortium and has received funding from the Euratom research and training program 2014-2018 and 2019-2020 under grant agreement No 633053. The views and opinions expressed herein do not necessarily reflect those of the European Commission.
 
Some of the simulations were performed on the Marconi-Fusion supercomputer located at CINECA, Italy.

The authors would like to thank E.~Strumberger, Q.~Yu, V.~Bandaru, E.~Viezzer and F.~Wieschollek for fruitful discussions.
 
\bibliographystyle{unsrtnat}
\bibliography{Meshcheriakov_RMPAUG}

\appendix

\chapter{Normalization}

We list a few normalization parameters for our simulations for reference. The normalization is described, e.g., in Ref.~\cite{Hoelzl_PoP2012}. 

\begin{align}
    n_{e,0} &= 8\cdot10^{19}\;\mathrm{m/s} \\
    m_{ion} &= m_D \\
    \rho_0  &= 2.7\cdot10^{-7}\;\mathrm{kg/m^3} \\
    \sqrt{\mu_0\rho_0} &= 5.8\cdot10^{-7} \\
    \sqrt{\mu_0/\rho_0} &= 2.2 \\
    B_0 &= 1.9 T
\end{align}

\end{document}